\documentclass[journal]{IEEEtran}

\usepackage{cite}
\usepackage{url}

\usepackage{graphicx}
\usepackage[cmex10]{amsmath}
\usepackage{array}
\usepackage{subfig}
\usepackage{booktabs}
\usepackage{here}

\def\Vec#1{\mbox{\boldmath $#1$}}
\def\figref#1{Fig.~\ref{#1}}
\def\figsref#1#2{Figs.~\ref{#1} and \ref{#2}}

\def\tabref#1{Table~\ref{#1}}
\def\secref#1{Section~\ref{#1}}
\def\eqref#1{(\ref{#1})}

\begin{document}

\title{Mosaicked multispectral image compression based on inter- and intra-band correlation}

\author{\authorblockN{
Kazuma Shinoda\authorrefmark{1}, Madoka Hasegawa\authorrefmark{1}, 
Masahiro Yamaguchi\authorrefmark{2},
Antonio Ortega\authorrefmark{3}} \\
\authorblockA{\authorrefmark{1}Graduate School of Engineering,
 Utsunomiya University, Utsunomiya, Japan} \\
\authorblockA{\authorrefmark{2}Global Scientific Information and Computing Center,
 Tokyo Institute of Technology, Yokohama, Japan} \\
\authorblockA{\authorrefmark{3}Signal and Image Processing Institute,
 University of Southern California, California, USA}
}


\maketitle

\begin{abstract}
Multispectral imaging has been utilized in many fields, but the cost of capturing and storing image data is still high.
Single-sensor cameras with multispectral filter arrays can reduce the cost of capturing images at the expense of slightly lower image quality.
When multispectral filter arrays are used, conventional multispectral image compression methods can be applied after interpolation, but the compressed image data after interpolation has some redundancy because the interpolated data are computed from the captured raw data.
In this paper, we propose an efficient image compression method for single-sensor multispectral cameras.
The proposed method encodes the captured multispectral data before interpolation.
We also propose a new spectral transform method for the compression of mosaicked multispectral images.
This transform is designed by considering the filter arrangement and the spectral sensitivities of a multispectral filter array.
The experimental results show that the proposed method achieves a higher peak signal-to-noise ratio at higher bit rates than a conventional compression method that encodes a multispectral image after interpolation, e.g.,
 3-dB gain over conventional compression when coding at rates of over 0.1 bit/pixel/bands.
\end{abstract}



\section{Introduction} \label{sec:intro}

Multispectral images (MSIs) are becoming increasingly important for a large number of applications, such as remote sensing, medical imaging, and digital archiving.
MSIs are images in which each pixel has multiple channels that carry information about its spectral content.
Multiband sensors with more than three channels have been used in remote sensing applications for many years.
Recently, several studies have reported an interest in the use of MSIs in the visible range of the spectrum in order to improve color reproduction \cite{ref:PDBurns1996,ref:NV_Yamaguchi2008}.

For the practical use of a spectrum-based imaging system,
 it is crucial to increase the number of bands in the image input device.
Various devices have been proposed for the acquisition of MSIs,
 such as a monochrome camera with a rotating filter wheel \cite{ref:MYamaguchi1997,ref:16bandsCamera2002}, a grating-prism (grism) \cite{ref:THyvarinen1998}, or a liquid-crystal tunable filter \cite{ref:RLevenson2003}.
These devices have certain limitations because of the complexity of assembling one or more prisms and multiple sensor arrays to detect signals.
To achieve an efficient solution for multispectral imaging, multispectral filter arrays (MSFAs) have been studied,
 inspired by the application of color filter arrays (CFA) in commercial digital RGB cameras.
Some studies have focused on filter array design and interpolation methods for MSFAs \cite{ref:KHirakawa2007,ref:LMiao2006_01,ref:YMonno2012_2,ref:KShinoda2013,ref:JBrauers2006,ref:LCondat2009}.

\begin{figure}[!t]
 \centerline{
  \subfloat[]{\includegraphics[width=1\linewidth]{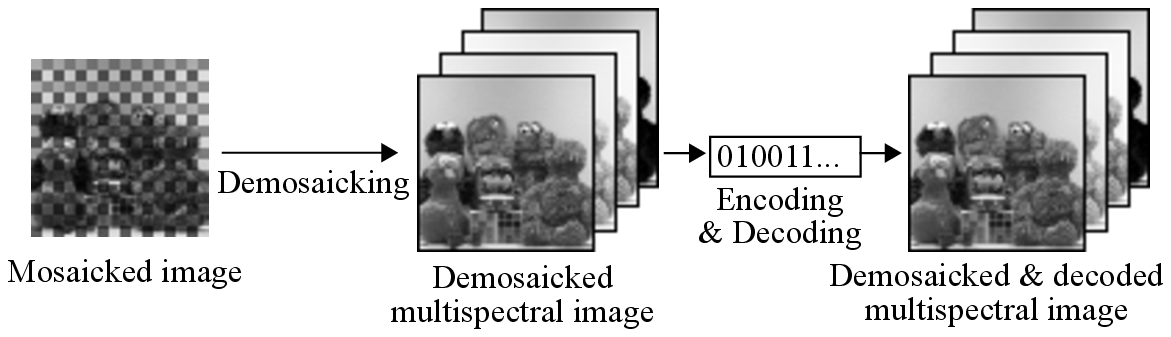}%
  \label{fig:EAI}}
 }
 \centerline{
  \subfloat[]{\includegraphics[width=1\linewidth]{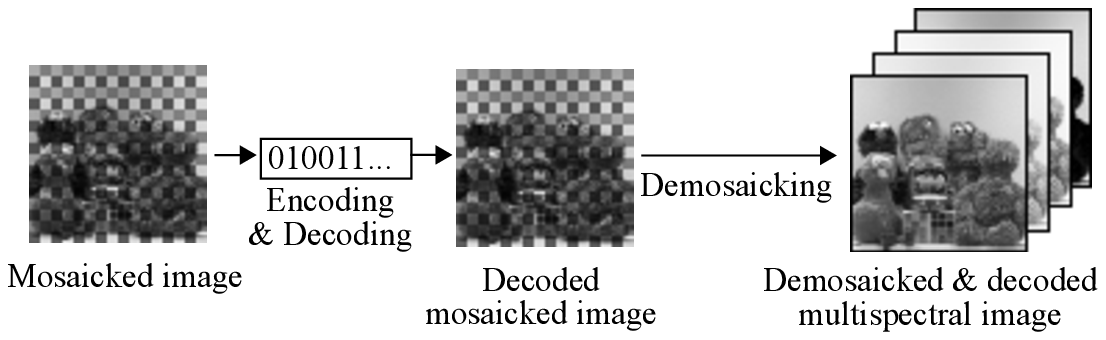}%
  \label{fig:EBI}}
 }
 \caption{Coding and demosaicking flow. (a) EAI (b) EBI.}
 \label{fig:EAIandEBI}
\end{figure}

Compression of mosaicked MSIs is another major challenge for practical use.
First, let us consider the compression of a mosaicked RGB image based on CFA.
In the context of CFA, most of the widely used compression methods such as JPEG and JPEG2000 are applied to demosaicked images.
\figref{fig:EAI} shows the overall diagram of such an imaging system, which we call ``encoding after interpolation'' (EAI) in this paper.
In the EAI chain, compression is employed to encode a full-resolution image after interpolation.
As an alternative, the imaging system shown in \figref{fig:EBI}, called ``encoding before interpolation'' (EBI), has also been proposed \cite{ref:KHChung2007,ref:YTTsai1991,ref:SYLee2001,ref:CCKoh2003,ref:BParrein2004,ref:XXie2005,ref:NXLian2006,ref:SYLee2009,ref:HSMalvar2012}.
In the EBI chain, the mosaicked image is encoded directly and demosaicking is employed after decompression.
EBI has the potential to achieve significant compression gains because the EBI approach encodes only 1/3rd of the amount of data encoded in EAI.

The concept of EBI encoding a mosaicked image before interpolation, can be applied irrespective of the number of bands.
Assuming $N$ to be the number of bands (in this paper, ``band'' or ``component'' means a plane number along optical spectral direction, and ``wavelength'' means the center wavelength of the spectral sensitivity of each color filter), we find that the EBI approach for MSI encodes only $1/N th$ of the amount of data encoded in EAI, but its compression method has not been studied thus far.
JPEG2000 Part 2 \cite{ref:JPEG2000part2}, 3D-SPIHT \cite{ref:3D-SPIHT-RARS}, and others \cite{ref:1DKLT2DDWT,ref:Joensuu2000,ref:Joensuu2001,ref:HSI2,ref:MSIC_Shinoda_JEI2011} support the compression for MSIs by treating them as volume data, but these compression methods have not been considered for the mosaicked image obtained from the EBI approach.
The compression methods for a mosaicked RGB image \cite{ref:KHChung2007,ref:YTTsai1991,ref:SYLee2001,ref:CCKoh2003,ref:BParrein2004,ref:XXie2005,ref:NXLian2006,ref:SYLee2009,ref:HSMalvar2012} cannot be applied directly to a mosaicked MSI because the number and arrangement of color filters are different between CFA and MSFA.
Moreover, the choice of the optical sensitivity of each color filter varies depending on the application.
Thus, reducing the redundancy of the mosaicked MSI based on the arrangement and optical sensitivity of a given MSFA is an important issue for EBI.

We propose a new coding method for EBI that considers the redundancy of the mosaicked MSI.
As a first step, the proposed method generates a sub-image from the mosaicked MSI by gathering the pixels from the same band to increase the intra-band correlation.
By rearranging all pixels of the mosaicked MSI into sub-images, we generate a pseudo-MSI.
Then, the spectral redundancy of the pseudo-MSI is reduced by one-dimensional (1-D) spectral transform, which corresponds to the multi-component transform (MCT) of JPEG2000.
Finally, the transformed image is encoded using JPEG2000.

In the proposed flow, the spectral transform is important.
The inter-band correlation of the pseudo-MSI is not constant along the band index because it depends on both the pattern and the optical sensitivity of an MSFA.
The redundancy of the inter-band correlation cannot be efficiently reduced by a simple differential pulse code modulation (DPCM) or discrete wavelet transform (DWT); instead, we first propose the use of Karhunen-Loeve transform (KLT) for the spectral transform along the band index.
A different KLT is computed for each image and then sent to the transform matrix as an MCT of JPEG2000.
This method has been proposed for full MSIs and has efficiently reduced redundancy \cite{ref:1DKLT2DDWT}; therefore, KLT is believed to perform efficiently for both pseudo-MSIs and full MSIs.

As an alternative to the data-dependent KLT, we propose a fixed transform for the spectral transform based on the MSFA information.
This is because the calculation cost of the covariance matrix for KLT is not negligible in a single-shot camera equipped with MSFA.
The proposed fixed transform matrix is generated using only the arrangement pattern and the center wavelengths of the MSFA filters.
Thus, once MSFA is generated for a certain camera, the transform matrix does not have to be re-calculated for each captured image.
This fixed transform achieves compression efficiency with lower computation cost and is particularly applicable to fast and low-cost MSFA-based cameras.

There are two contributions of our study.
First, we analyze the performance advantages of EBI for MSI, which have thus far not been studied in detail.
Second, we propose two spectral transforms to reduce the redundancy of mosaicked MSI.
Our previous work \cite{ref:KShinoda_PCS2013} proposed the base algorithm of EBI for MSI, but the validity of the fixed transform had not been verified.
In this study, the assumed correlation coefficients in the fixed transform are compared with those generated from a real image in order to validate the fixed transform.
Further, we examine the coding performance in various situations by using not only natural images but also a vegetation image in order to show the robustness of the proposed method.

The rest of this paper is organized as follows:
First, conventional compression methods for MSI are introduced in \secref{sec:MSIC};
then, the compression method for mosaicked RGB images is introduced in \secref{sec:CFA}.
The proposed compression method is presented in \secref{sec:EBICodingMethod}.
The results are given in \secref{sec:Experiment}.
In \secref{sec:Discussion}, we compare the coding performance when using different parameters with a vegetation image.
In \secref{sec:Conclusion}, we conclude this paper.

\section{Multispectral image compression} \label{sec:MSIC}

\begin{figure}[!t]
 \centerline{
  \subfloat[]{\includegraphics[width=0.5\linewidth]{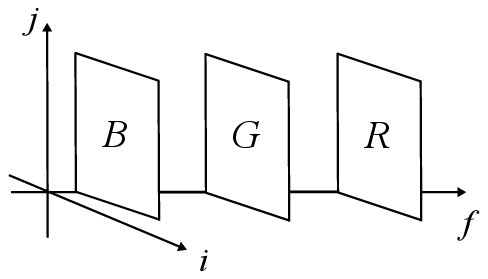}%
  \label{fig:RGBintervals}}
 \hfill
  \subfloat[]{\includegraphics[width=0.5\linewidth]{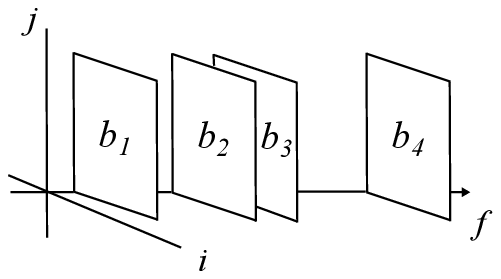}%
  \label{fig:MSIintervals}}
 }
 \caption{Example of spectral intervals. Here, $i$ and $j$ denote the spatial directions, and $f$ represents the optical wavelength. (a) RGB (b) 4-band MSI $\{b_{1},b_{2},b_{3},b_{4}\}$.}
 \label{fig:RGBandMSIintervals}
\end{figure}

MSIs can use different sets of bands depending on the specific application, and the reduction of redundancy along the optical spectral direction is an important factor for compression.
In some applications, the spectral intervals of MSI may not be constant as shown in \figref{fig:RGBandMSIintervals}, and the inter-band correlations are different from those in cases of RGB.
Several techniques have been proposed in the literature that exploit the spectral correlation characteristic of MSI data.
In early publications \cite{ref:PJReady1973,ref:GFernandez1996}, KLT was applied to reduce the redundancy across the bands.
In \cite{ref:1DKLT2DDWT}, KLT was used as a 1-D spectral transform, following which a two-dimensional (2-D) wavelet transform was used as the spatial compression scheme.
The separation of the three-dimensional (3-D) wavelet transform into spectral and spatial domains has been examined by Kaarna and Parkkinen\cite{ref:Joensuu2000,ref:Joensuu2001}.
For remote sensing, coding methods with 1-D spectral and 2-D spatial transforms have been applied to MSI\cite{ref:FGVilchez2009,ref:HSI2}.
As a state-of-the-art compression method, Part 2 of JPEG2000 \cite{ref:JPEG2000part2} makes provisions for arbitrary MSI.
The pixel structure of MSI is almost equivalent to that of RGB except for the number of components; therefore, the baseline algorithm using DWT and embedded block coding with optimized truncation (EBCOT) can be applied to MSI by specifying the MCT instead of $\rm{YC_bC_r}$.
In particular, applying KLT as MCT leads to a considerably better rate distortion performance than applying other transforms such as DWTs \cite{ref:QDu2007}.
Therefore, coding approaches based on 1-D spectral transform and 2-D spatial transform are popular choices because of their coding performance and their compatibility with JPEG2000.

The inter-band correlation should also be considered in the case of compression for mosaicked MSI.
In a mosaicked image, spectral redundancy exists between neighboring pixels in the same plane.
Thus, not only intra-band correlation but also inter-band correlation should be considered when reducing the redundancy between neighboring pixels.
However, thus far, a concrete method to reduce such redundancy for mosaicked MSI has not been studied.
Separating spectral and spatial transforms and considering the inter-band correlation would lead to better performance in many MSI applications.
Moreover, separating these transforms can help to maintain the compatibility with JPEG2000.


\section{Image compression for mosaicked RGB image} \label{sec:CFA}

\begin{figure}[!t]
 \begin{center}
  \includegraphics[width = 1\linewidth]{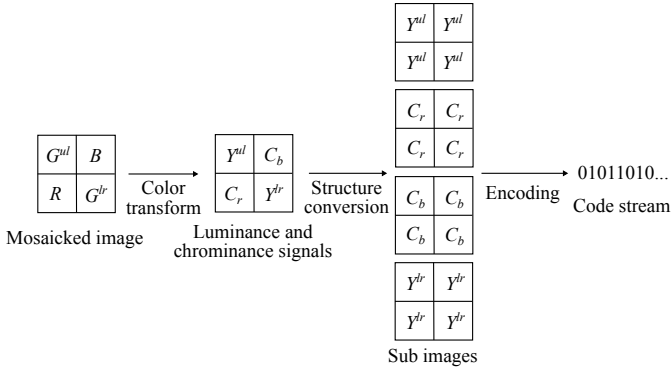}
 \end{center}
 \caption{Conventional EBI flow. The superscripts ul and ur indicate the upper left and lower right positions in a 2 $\times$ 2 CFA block.}
 \label{fig:EBIforRGBFlow}
\end{figure}

As simple lossy EBI compression methods can be developed by considering the mosaicked image as a grayscale image, compression techniques can be applied directly to the mosaicked images, ignoring pixel color labels \cite{ref:YTTsai1991}.
However, this method often leads to poor compression performance because of the artificial discontinuities due to the interlaced color components.
In general, color transform and mosaicked structure conversion should be used prior to image compression in order to achieve better performance \cite{ref:SYLee2001,ref:CCKoh2003,ref:SYLee2009,ref:HSMalvar2012,ref:NXLian2006}, as shown in \figref{fig:EBIforRGBFlow}.
The structure conversion step transforms the mosaicked pixels corresponding to the same color filters into a structure more appropriate for image coding.
This process creates sub-images that contain more natural edges than a mosaicked image.
A color transform to decorrelate the color components can be used for removing artificial discontinuities in mosaicked images.
In the above-mentioned methods, first, the four color values in each $2 \times 2$ Bayer unit are converted to two luminance values and two chrominance values.
The resulting chrominance values reside in a rectangular lattice with a size four times smaller than the size of
CFA, while the luminance values populate a quincunx lattice that is half the size of CFA.
Each chrominance plane can be compressed using standard techniques after converting to a rectangular lattice, whereas the luminance plane cannot be directly compressed by standard techniques because it consists of a quincunx lattice.
Ko et al. \cite{ref:CCKoh2003} and Malvar et al. \cite{ref:HSMalvar2012} separate the luminance signal into odd and even indices, and then, these two sub-images are compressed using standard techniques such as JPEG or JPEG-XR.
Lee et al. \cite{ref:SYLee2009} transform the luminance signal into a rectangular lattice by shift and rotation.

These studies suggest that structure conversion is an efficient approach for a mosaicked image because the converted image contains more natural edges than the mosaicked image.
Intra-band coding by using standard compression techniques can be performed efficiently on the converted image.
Structure conversion is also considered an efficient approach in the case of MSI.
The conventional EBI approach for CFA cannot be directly applied to the case of MSI because the filter pattern is different, but the concept of merging same band pixels into a sub-image can be exploited to improve the compression efficiency.

The performance difference between EAI and EBI in the previous works is also a notable point for extending to MSI case.
EBI can definitely reduce the data size of lossless compression as compared to EAI because the information of a mosaicked image is completely included in a demosaicked image on the EAI \cite{ref:KHChung2007,ref:HSMalvar2012}.
On the other hand, it has been shown that the lossy compression performance of EBI is not always superior to that of EAI, leading to a performance trade-off \cite{ref:SYLee2001,ref:CCKoh2003,ref:SYLee2009,ref:HSMalvar2012,ref:NXLian2006}.
In EBI, the total error between the original image and a decoded full-resolution image includes both the demosaicking and the compression error.
In general, EBI outperforms EAI at higher bit rates because mosaicked images compressed in EBI have fewer pixels.
EAI outperforms EBI at lower bit rates because full-resolution images have larger intra-band and inter-band correlations than the corresponding mosaicked images.
Although EAI has been used widely in recent imaging systems, most of the studies show that EBI is superior to EAI at higher bit rates.

In the case of MSI, almost the same trend can be seen as in the case of RGB because the verification models in these previous works are consistent with the case of MSI except for the number of bands.
However, the performance change from RGB to MSI should be verified under various experimental conditions.


\section{Image compression for mosaicked MSI} \label{sec:EBICodingMethod}

EBI techniques for RGB cannot be applied directly to MSI because the algorithms are specific to Bayer CFA.
The G channel of the mosaicked image is encoded differently from the R and B channels because the sample density is different in each channel.
In contrast, MSFAs have more bands and different patterns than CFAs.
An existing MSFA-based camera has a certain MSFA pattern, but there are many possible MSFA configurations depending on the future application.
EBIs for MSI require flexibility to be able to reduce the redundancy of the mosaicked pixels under a given MSFA irrespective of the number of bands and the patterns.

\begin{figure}[!t]
 \centerline{
  \subfloat[]{\includegraphics[width=1\linewidth]{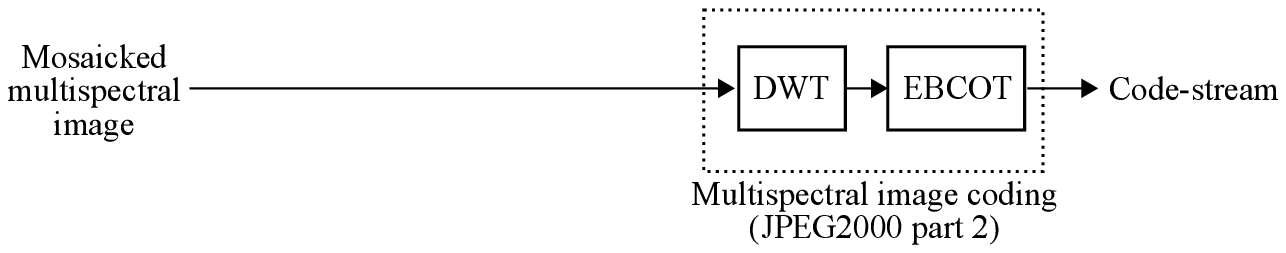}%
  \label{fig:CodingFlowPC}}
 }
 \centerline{
  \subfloat[]{\includegraphics[width=1\linewidth]{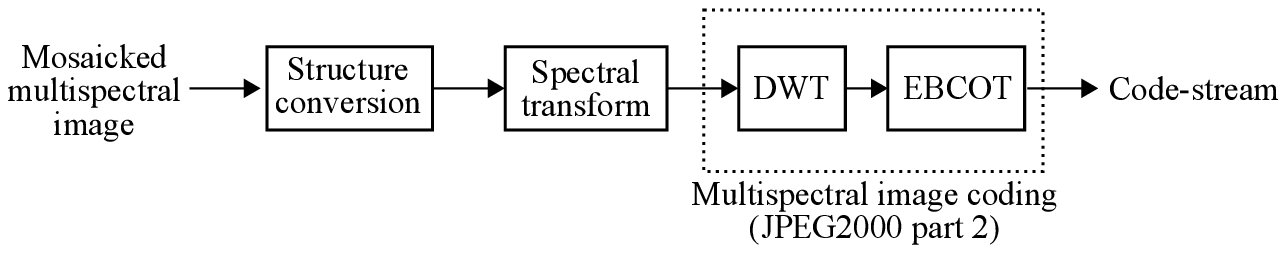}%
  \label{fig:CodingFlowPro}}
 }
 \caption{Coding flow for mosaicked MSI in EBI. (a) Direct coding (b) EBI with KLT and EBI with fixed transform.}
 \label{fig:CodingFlow}
\end{figure}

Further, the optical wavelength intervals of MSFA should be considered because the wavelength intervals of the filters are not the same.
Many EBIs for RGB calculate $C_{b}$ and $C_{r}$, but a similar conversion is not possible for MSI.
Only subtraction between bands is not sufficient to reduce the spectral redundancy.
We propose an efficient transform for reducing the redundancy that is based on both 1-D spectral and 2-D spatial correlation coefficients.

In this section, we propose a simple method where {\it direct coding} is used first, and then, two compression methods, namely {\it EBI with KLT} and {\it EBI with fixed transform}, are proposed.
\figref{fig:CodingFlow} shows the overview of the coding flow, wherein we use JPEG2000. 
{\it Direct coding} is a simple compression method that directly encodes an image by using JPEG2000, as shown in \figref{fig:CodingFlowPC}. 
{\it EBI with KLT} and {\it EBI with fixed transform} follow the same flow as that shown in \figref{fig:CodingFlowPro}. The main difference with respect to direct coding (\figref{fig:CodingFlowPC}) is the introduction of the {\it structure conversion} and {\it spectral transform} parts for reducing spectral redundancy, and then, the use of JPEG2000 part 2 for MSI coding.

\subsection{Direct coding}
Mosaicked MSIs can be regarded as grayscale images whose pixels correspond to only one band.
Therefore, the mosaicked image can be encoded directly by using JPEG2000.
We call this approach ``Direct coding'' in this paper.
Direct coding often leads to poor compression performance because of the discontinuities after mosaicking \cite{ref:YTTsai1991}.
Further, by comparing the mosaicked images obtained in the RGB and MSI cases, we find that the observed discontinuities are slightly different, because the wavelength intervals between bands are different.
We evaluate the performance of direct coding in the next section.

\subsection{Proposed EBI with KLT} \label{ssec:KLT}

\begin{figure}[!t]
 \begin{center}
  \includegraphics[width = 0.7\linewidth]{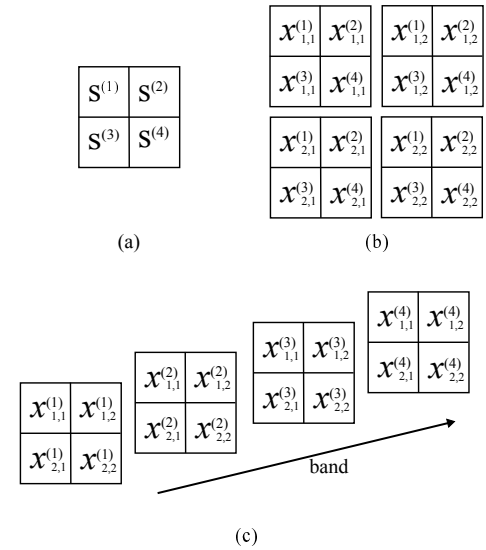}
 \end{center}
 \vspace{-3mm}
 \caption{Structure conversion example on EBI. (a) MSFA (b) Captured $4 \times 4$ image with $2 \times 2$ MSFAs (c) Converted $2 \times 2 \times 4$ image.}
 \label{fig:ReshapingMethod}
\end{figure}

A more efficient EBI approach is to map a mosaicked image to a series of smaller images.
All pixels corresponding to a given band are placed in a 2-D image on the 2-D plane; thus, after remapping, we have a pseudo-MSI with smaller planes, each containing all the samples at that spectral component.
The generated pseudo-MSI can be encoded by JPEG2000 with MCT.
In the following paragraphs, structure conversion methods and spectral transforms suitable for mosaicked MSI are proposed assuming that we use JPEG2000.

The proposed EBI consists of three parts: (i) {\it Structure conversion}, where an $N$-band pseudo-MSI is formed with $1/N$-resolution planes from the mosaicked image, (ii) {\it Spectral transform}, where a 1-D spectral transform is used across spectral bands in the $N$-band MSI in order to exploit their redundancy, and (iii) {\it Coding}, where a conventional coding scheme such as JPEG2000 is applied spatially to each band after it has been processed with the 1-D spectral transform.

An example of a 4-band MSFA and a captured $4 \times 4$ image with $2 \times 2$ MSFAs are shown in Figs.~\ref{fig:ReshapingMethod}a and b, respectively.
Let ${\rm S}^{(n)} (n = 1, 2, \ldots, N)$ be a filter, where a smaller $n$ value indicates a shorter wavelength.
Note that ${\rm S}$ is just a label to distinguish filters (it is not a variable).
$x^{(n)}_{i,j}$ denotes a captured signal at a block-based position $(i, j)$ in band $n$.
As mentioned before, the mosaicked image in \figref{fig:ReshapingMethod}b can be encoded as a grayscale image because the image has only one channel for each pixel.
However, we can see that the neighboring pixels (e.g., $x^{(1)}_{1,1}$ and $x^{(2)}_{1,1}$ in \figref{fig:ReshapingMethod}b) correspond to different bands and that pixels captured from the same band ($x^{(1)}_{1,1}$ and $x^{(1)}_{1,2}$ in \figref{fig:ReshapingMethod}b) are located away from each other.
In order to improve the coding performance, all pixels corresponding to the same band (e.g., $x^{(1)}_{1,1}$ and $x^{(1)}_{1,2}$) are placed in a 2-D image on the 2-D plane. This is the {\it structure conversion} process that yields $N$ images, each containing $1/N$ of the pixels in the original mosaicked image.
The {\it structure conversion} process in a 4-band MSFA is shown in \figref{fig:ReshapingMethod}c.
These new images are arranged in the ascending order of $n$.

In typical MSIs (not obtained using an MSFA), there exists significant correlation across bands, which is often exploited by using a 1-D spectral transform before spatial coding.
However, due to the original image being mosaicked in the MSFA imaging system, there is information about {\em only one band} at each spatial location in \figref{fig:ReshapingMethod}b.
Thus, although the image of \figref{fig:ReshapingMethod}c seems like a normal MSI at the first glance, the pixels in the same spatial position ($x^{(1)}_{i,j}, x^{(2)}_{i,j}, \ldots, x^{(N)}_{i,j}$) are not located in the same position in \figref{fig:ReshapingMethod}b.
The correlation across the bands is low compared to that of a true MSI, but not zero.
The {\it spectral transform} in the proposed system aims to exploit the redundancy across the bands in \figref{fig:ReshapingMethod}c, while taking into account the effect of mosaicking.
We assume that there will be more high-frequency information in the spectral direction than for a true MSI.
Therefore, KLT would be an appropriate method for exploiting the redundancy as it can adapt to the specific correlation in 1-D signals.

We define the converted signal $\Vec{x}_{i,j}$ as $[x^{(1)}_{i,j} x^{(2)}_{i,j} \ldots x^{(N)}_{i,j}]^{T}$,
 and the transformed signal $\Vec{y}_{i,j}$ as $[y^{(1)}_{i,j} y^{(2)}_{i,j} \ldots y^{(N)}_{i,j}]^{T}$.
$\Vec{R}_{x}$ is calculated using the singular value decomposition (SVD) of $\Vec{x}_{i,j}$.
Let $\Vec{T}_{k}$ be the matrix of eigenvectors of the correlation matrix $\Vec{R}_{x}$.
The eigenvectors are lined row-by-row on $\Vec{T}_{k}$.
A transformed coefficient is calculated as follows:
\begin{eqnarray}
 \Vec{y}_{i,j} = \Vec{T}_{k} \Vec{x}_{i,j}. \label{eq:KLTTransform}
\end{eqnarray}
After 1-D KLT, $\Vec{y}$ is encoded using JPEG2000 \cite{ref:JPEG2000part2}.

\subsection{Proposed EBI with fixed transform} \label{ssec:Fixed}
KLT based on the captured data would be an appropriate method for exploiting redundancy as it can adapt to the specific correlation in 1-D signals.
However, the transform matrix would need to be updated on the basis of the input data, which may be very costly.
Alternatively, it is questionable whether a training dataset would be reliable, as the specific spatial correlation of the images can play a significant role in the transform.
Instead of the KLT discussed in \secref{ssec:KLT}, we propose a fixed transform design that models the spectral and spatial correlation based on simple assumptions.
We expect this to be a more robust solution, which does not require retraining for specific content.
Note that (i) the {\it structure conversion} and (iii) {\it coding} parts are the same processes as those discussed in \secref{ssec:KLT}.

\begin{figure}[!t]
 \begin{center}
  \includegraphics[width = 1\linewidth]{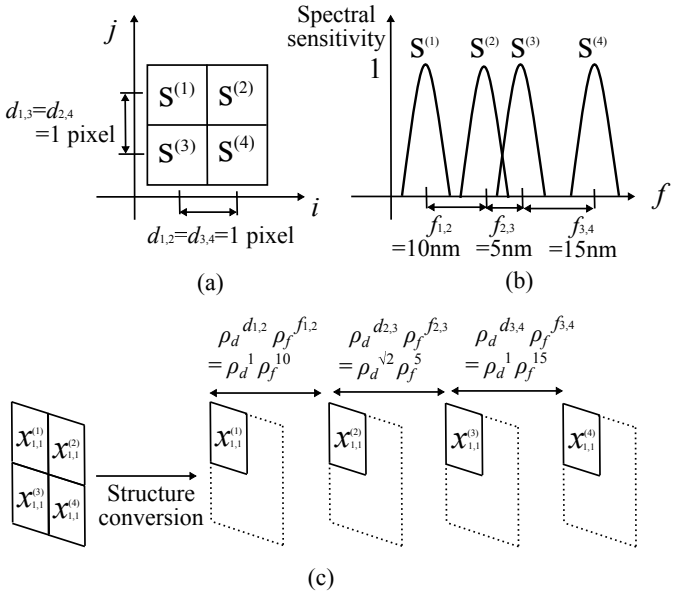}
 \end{center}
 \vspace{-3mm}
 \caption{Example of correlation coefficients in the spectral and spatial directions in MSFA. (a) Spatial correlation of MSFA (b) Spectral correlation of MSFA (c) Correlation coefficients of converted mosaicked MSI.}
 \label{fig:Correlation}
\end{figure}

We assume a simple relation between the captured pixels on the basis of the MSFA conditions.
\figref{fig:Correlation} shows an example of the spectral and spatial correlations of $2\times2$ MSFA.
\figref{fig:Correlation}a shows an example of $2\times2$ MSFA; \figref{fig:Correlation}b, the optical spectral sensitivity of each color filter in front of the sensor; and \figref{fig:Correlation}c, a captured mosaicked image and a converted one.
Here, $d_{m, n}$ denotes the Euclidean distance (in pixels) between the filters ${\rm S}^{(m)}$ and ${\rm S}^{(n)}$ on MSFA, $f_{m, n}$ represents the difference in the center wavelengths (in nanometers), $\rho_{_{d}}$ refers to the spatial correlation coefficient per pixel, and $\rho_{_{f}}$ denotes the spectral correlation coefficient per nanometer.
$f_{m, n}$ and $d_{m, n}$ are automatically determined after the specification of the MSFA pattern and the spectral sensitivity.

In this paper, $\rho_{_{f}}$ and $\rho_{_{d}}$ are assumed to be constant values given in advance.
$\rho_{_{d}}$ can be set by assuming that the signal is a first-order Markov process with the correlation coefficient $\rho_{_{d}} = 0.95$\cite{ref:RJClarke1981,ref:Spr2009}, which is widely used in the field of image processing.
In case that a captured object is limited, as in medical applications, another effective way to obtain $\rho_{_{d}}$ more properly is to capture the target objects by using a multispectral or RGB camera in advance and then, compute $\rho_{_{d}}$ from the captured image.
Similarly, $\rho_{_{d}}$ and $\rho_{_{f}}$ should be obtained from real data in advance, but doing so is very difficult as a spectrometer or multi-shot multispectral camera is required for measuring real spectral data.
According to the examination of \cite{ref:WKPratt1976}, the correlation coefficients of real spectral reflectance can be expressed by a first-order Markov model as $0.95 < \rho_{_{f}} < 1$, and $\rho_{_{f}}$ is approximately equal to 0.995 per nanometer \cite{ref:Murakami2009}. The
$\rho_{_{f}}$ and $\rho_{_{d}}$ values can be assumed to be 0.95 and 0.995, respectively, if it is difficult to obtain the real correlation coefficients; we use these values in this paper.
More details about these values are provided in \secref{sec:Discussion}.

The actual correlation coefficient between ${\rm S}^{(m)}$ and ${\rm S}^{(n)}$ after structure conversion can be calculated from the captured pixels, but the calculation cost is not negligible as mentioned before.
The goal of the fixed transform is to obtain the correlation matrix of the converted image by using only the information of the MSFA.
Here, consider the relation between band 1 and band 2 as an example.
$x^{(1)}_{1,1}$ is sampled with ${\rm S}^{(1)}$, and $x^{(2)}_{1,1}$ is sampled with ${\rm S}^{(2)}$.
The spatial distance between ${\rm S}^{(1)}$ and ${\rm S}^{(2)}$ is $d_{1, 2} = 1$ pixel in \figref{fig:Correlation}a.
${\rm S}^{(1)}$ and ${\rm S}^{(2)}$ have different optical sensitivity values, as shown in \figref{fig:Correlation}b, and the distance between the center wavelengths is $f_{1, 2} = 10$ nm.
The spatial correlation coefficient can be assumed to be $\rho_{_{d}}^{d_{m, n}}$ on the basis of the first-order Markov model, and the same applies to the spectral correlation.
The correlation coefficient between $x^{(1)}_{1,1}$ and $x^{(2)}_{1,1}$ is finally $\rho_{_{d}}^{d_{1, 2}} \cdot \rho_{_{f}}^{f_{1, 2}} = \rho_{_{d}}^{1} \cdot \rho_{_{f}}^{10}$, as shown in \figref{fig:Correlation}c.
The correlation coefficients of the other combinations can also be obtained by using the same procedure.
Finally, the fixed transform matrix is calculated from the SVD of the correlation coefficients.
Thus, the fixed transform matrix can be obtained if the arrangement pattern (\figref{fig:Correlation}a) and the optical sensitivity (\figref{fig:Correlation}b) of MSFA are given in advance.

In a general case, we define an $N \times N$ correlation matrix $\Vec{R}_{fd}$ calculated from a spectral correlation matrix $\Vec{R}_{f}$
 and the spatial correlation matrix $\Vec{R}_{d}$ of the ${\rm S}^{(n)}$.
First, to derive $\Vec{R}_{f}$, we denote the center wavelengths of $N$-band filters as $f_{1},$ $f_{2},$ $\ldots,$ $f_{N}$.
The difference in the wavelengths is $f_{1,2},$ $f_{1,3},$ $\ldots,$ $f_{1,N},$ $f_{2,3},$ $\ldots,$ $f_{N-1,N}$.
The spectral correlation matrix $\Vec{R}_{f}$ can be defined as follows:
\begin{eqnarray}
\Vec{R}_{f} &=&
\left[
  \begin{array}{ccccc}
 1 & \rho_{_{f}}^{f_{1,2}} & \rho_{_{f}}^{f_{1,3}} & \ldots & \rho_{_{f}}^{f_{1,N}} \\
 \rho_{_{f}}^{f_{1,2}} & 1 & \rho_{_{f}}^{f_{2,3}} & \ldots & \rho_{_{f}}^{f_{2,N}} \\
 \rho_{_{f}}^{f_{1,3}} & \rho_{_{f}}^{f_{2,3}} & 1 & \ldots & \rho_{_{f}}^{f_{3,N}} \\
 \vdots & \vdots & \vdots & \ddots & \vdots \\
 \rho_{_{f}}^{f_{1,N}} & \rho_{_{f}}^{f_{2,N}} & \rho_{_{f}}^{f_{3,N}} & \ldots & 1
  \end{array}
\right], \label{eq:Markov}
\end{eqnarray}
 where $\rho_{_{f}}$ indicates the correlation coefficient per nanometer spacing
 and $\rho_{_{f}}$ is raised to the power $f_{k,l}$.
This matrix can satisfactorily approximate the correlation matrix of the spectrum when $0.95 < \rho_{_{f}} < 1$\cite{ref:WKPratt1976}.

For calculating a spatial correlation matrix $\Vec{R}_{d}$,
 let the Euclidean distance between the color filters be $d_{1,2},$ $d_{1,3},$ $\ldots,$ $d_{1,N},$ $d_{2,3},$ $\ldots,$ $d_{N-1,N}$.
For example, we choose $d = 1$ if the spatial centers direct neighbors (N, W, E, S) in the 4-connected grid, and $d = \sqrt{2}$ if they are on the diagonals.
The spatial correlation matrix $\Vec{R}_{d}$ is then defined as follows:
\begin{eqnarray}
\Vec{R}_{d} &=&
\left[
  \begin{array}{ccccc}
 1 & \rho_{_{d}}^{d_{1,2}} & \rho_{_{d}}^{d_{1,3}} & \ldots & \rho_{_{d}}^{d_{1,N}} \\
 \rho_{_{d}}^{d_{1,2}} & 1 & \rho_{_{d}}^{d_{2,3}} & \ldots & \rho_{_{d}}^{d_{2,N}} \\
 \rho_{_{d}}^{d_{1,3}} & \rho_{_{d}}^{d_{2,3}} & 1 & \ldots & \rho_{_{d}}^{d_{3,N}} \\
 \vdots & \vdots & \vdots & \ddots & \vdots \\
 \rho_{_{d}}^{d_{1,N}} & \rho_{_{d}}^{d_{2,N}} & \rho_{_{d}}^{d_{3,N}} & \ldots & 1
  \end{array}
\right], \label{eq:Markov2}
\end{eqnarray}
 where $\rho_{_{d}}$ indicates the spatial correlation between neighboring pixels
 and $\rho_{_{d}}$ is raised to the power $d_{k,l}$.
Finally, the spatial-spectral correlation matrix $\Vec{R}_{fd}$ is calculated by
\begin{eqnarray}
 \Vec{R}_{fd} = \Vec{R}_{f} \circ \Vec{R}_{d}, \label{eq:CorrelationMatrix}
\end{eqnarray}
 where $\circ$ denotes the Hadamard product.
Let $\Vec{T}_{f}$ be a matrix of the eigenvectors of the correlation matrix
$\Vec{R}_{fd}$ calculated using the SVD.
The eigenvectors are lined row-by-row on $\Vec{T}_{f}$.
A transformed coefficient is calculated as follows:
\begin{eqnarray}
 \Vec{y}_{i,j} = \Vec{T}_{f} \Vec{x}_{i,j}, \label{eq:FixedMatrixTransform}
\end{eqnarray}
 where $\Vec{x}_{i,j} = [x^{^{(1)}}_{i,j}, x^{^{(2)}}_{i,j}, \ldots, x^{^{(N)}}_{i,j}]^T$ denotes the converted data
 and $\Vec{y}_{i,j} = [y^{^{(1)}}_{i,j}, y^{^{(2)}}_{i,j}, \ldots, y^{^{(N)}}_{i,j}]^T$ represents the transformed data.
After 1-D fixed transform, $\Vec{y}$ is encoded by JPEG2000 in the {\it coding} part.

\begin{figure}[!t]
 \centerline{
  \subfloat[]{\includegraphics[width=0.9\linewidth]{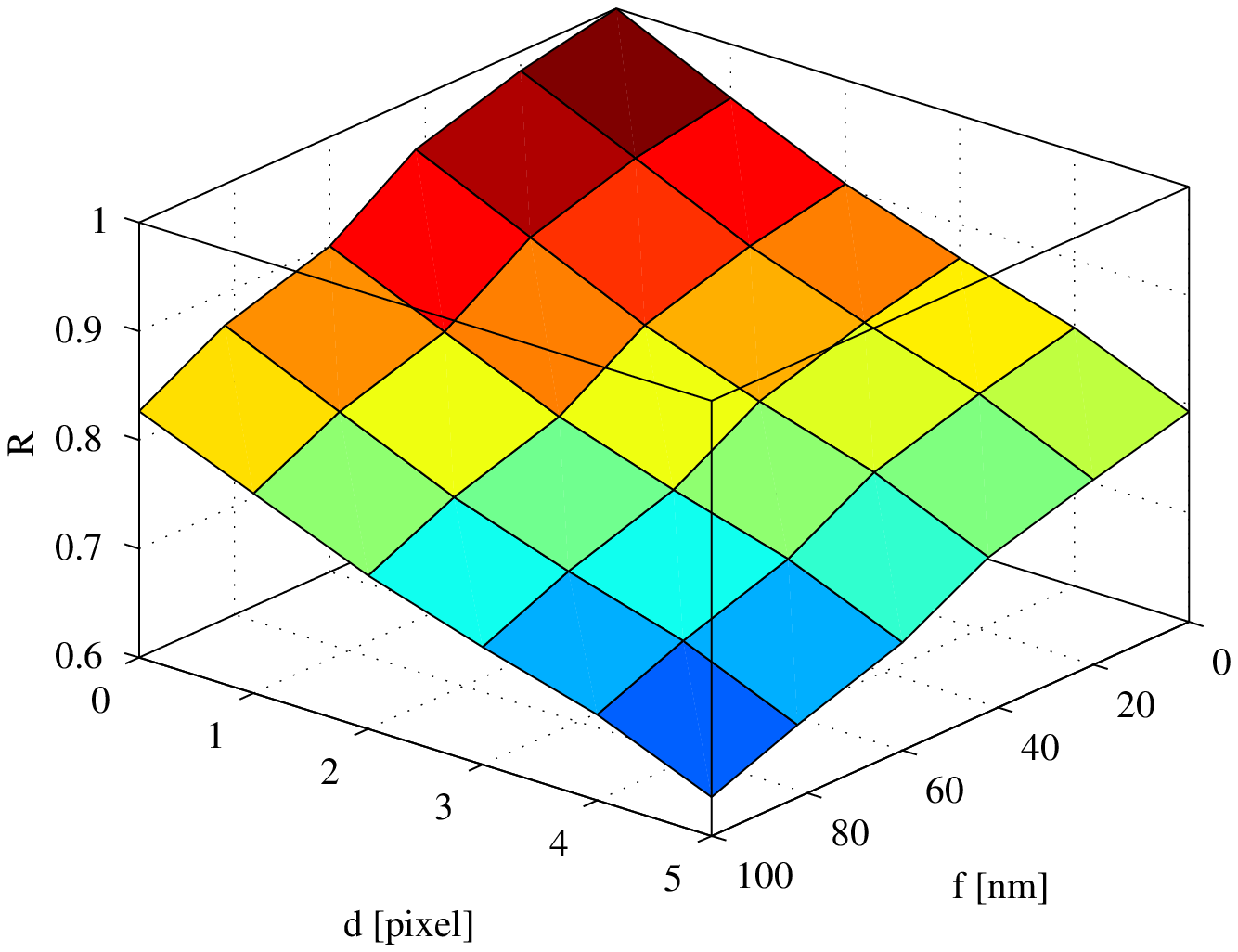}%
  \label{fig:CorrelationOfReal}}
 }
 \centerline{
  \subfloat[]{\includegraphics[width=0.9\linewidth]{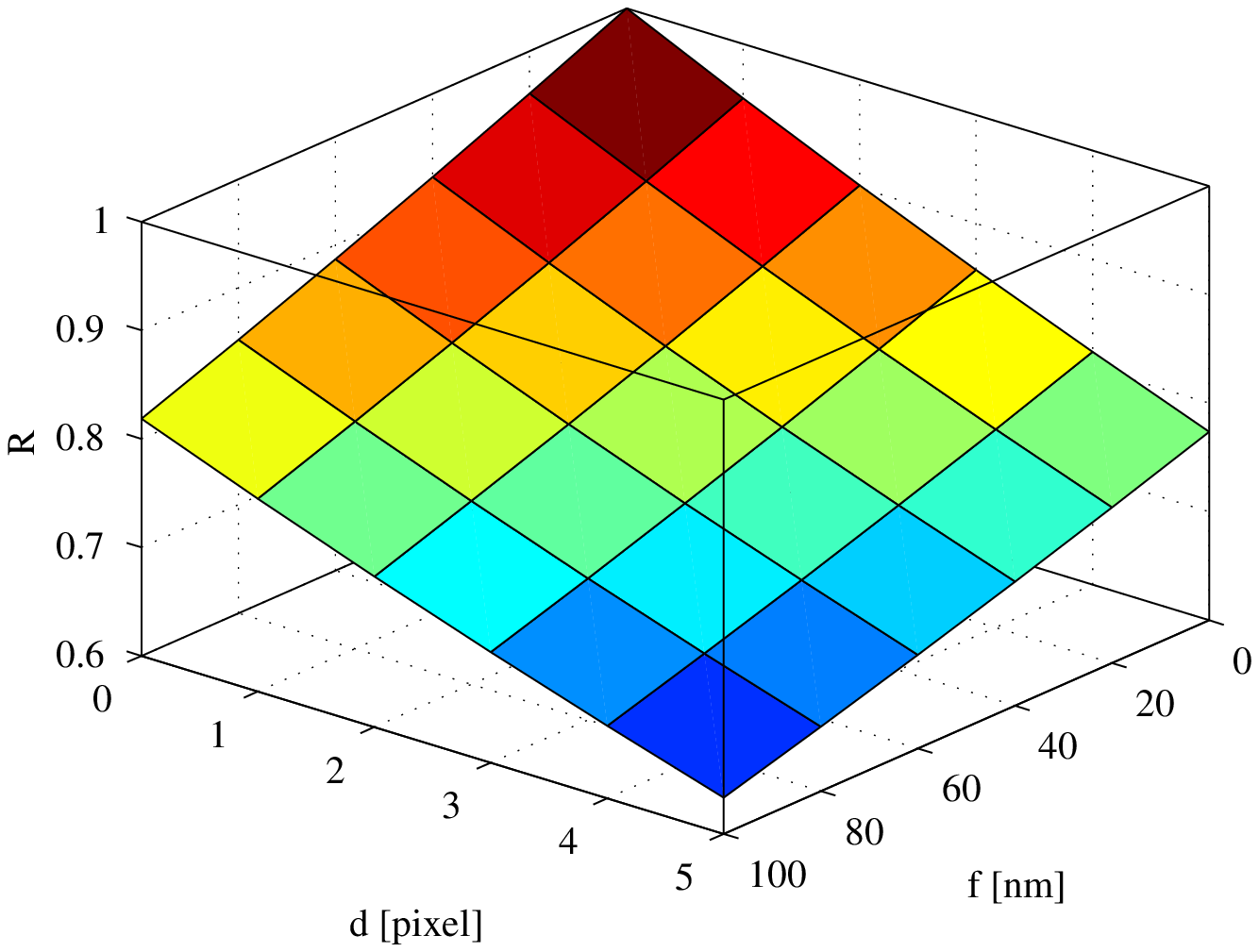}%
  \label{fig:Correlation OfFixed}}
 }
 \caption{Comparison of correlation coefficients. (a) $R$ calculated from a real MSI ({\it Toys}) (b) $R$ calculated from the proposed method and the spectral sensitivity of filters (\eqref{eq:FixedMatrixTransform} and \figref{fig:Filter16}). $\rho_{_{f}}$ and $\rho_{_{d}}$ are set to 0.995 and 0.95, respectively.}
 \label{fig:CompOfCorrelation}
\end{figure}

The fixed transform is attributed to the hypothesis of the Markov process, which states that the correlation coefficients between mosaicked pixels can be modeled by $\rho_{_{d}}$, $\rho_{_{f}}$, and the MSFA pattern.
For verifying the appropriateness of the assumed model, we compare the correlation coefficients obtained using \eqref{eq:CorrelationMatrix} and those calculated from a real MSI.
\figref{fig:TestImages} shows test images that consist of $512 \times 512$ pixels, 12 bits/pixel with 16 bands, and we use \figref{fig:Toys} for calculating the real correlation coefficients.
The images were captured with a full-resolution camera \cite{ref:16bandsCamera2002} equipped with 16-band filters, as shown in \figref{fig:Filter16}.
\figref{fig:CorrelationOfReal} shows the real correlation coefficients calculated from $Toys$, and \figref{fig:CompOfCorrelation} shows the proposed correlation model calculated from\eqref{eq:CorrelationMatrix}.
Here, $\rho_{_{f}}$ and $\rho_{_{d}}$ are set to 0.995 and 0.95, respectively.
As shown in \figref{fig:CompOfCorrelation}, the distribution of the proposed correlation coefficients is almost identical to that of the real coefficients.
The mean square error is 0.713, and the correlation is 0.941 between these two distributions.
From the perspective of an objective evaluation, the assumed correlation coefficients in the fixed transform are consistent with the real correlation coefficients.

\section{Experimental results} \label{sec:Experiment}

\begin{figure}[!t]
 \centerline{
  \subfloat[]{\includegraphics[width = 0.3\linewidth]{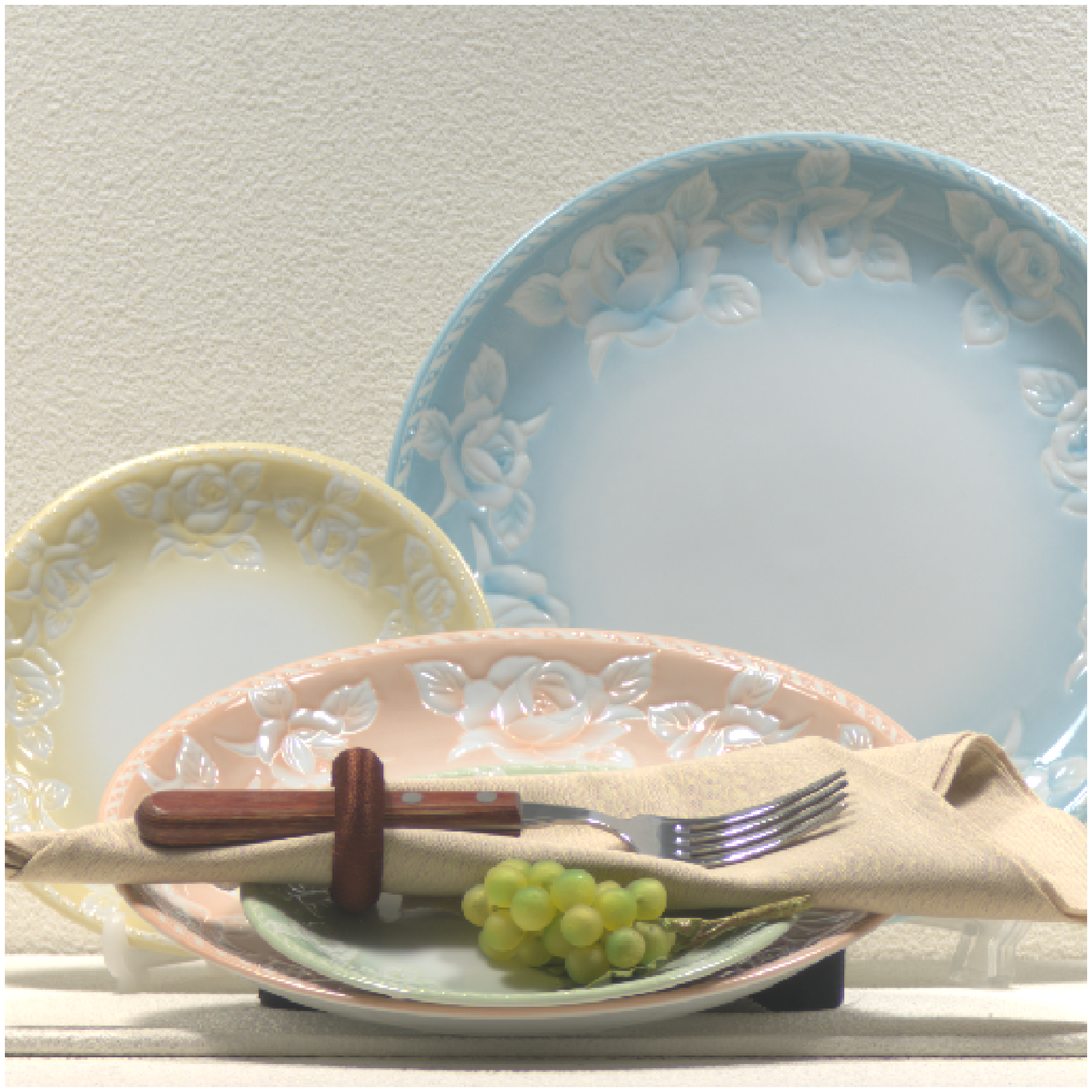}%
  \label{fig:Dishes}}
  \hfil
  \subfloat[]{\includegraphics[width = 0.3\linewidth]{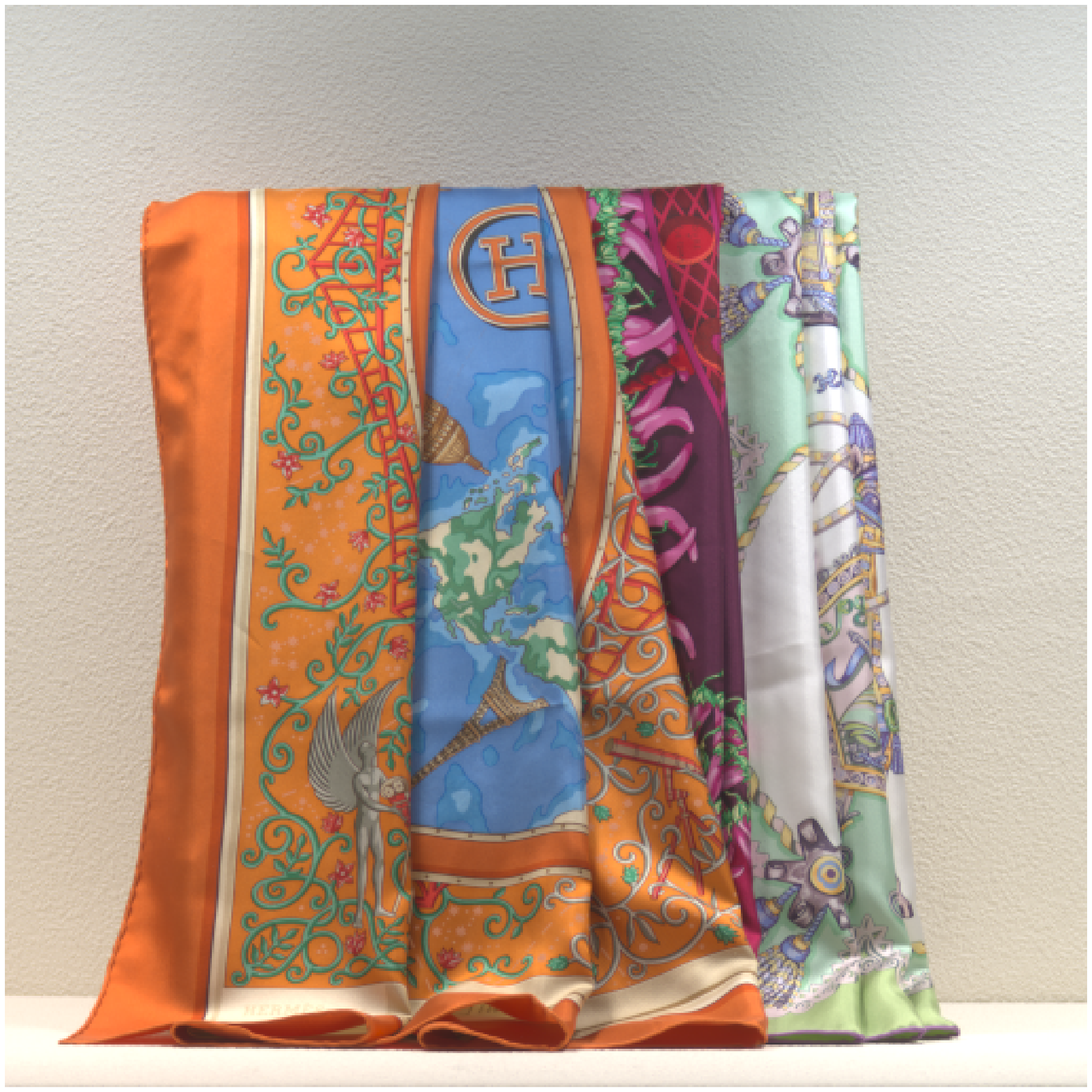}%
  \label{fig:Scarf}}
  \hfil
  \subfloat[]{\includegraphics[width = 0.3\linewidth]{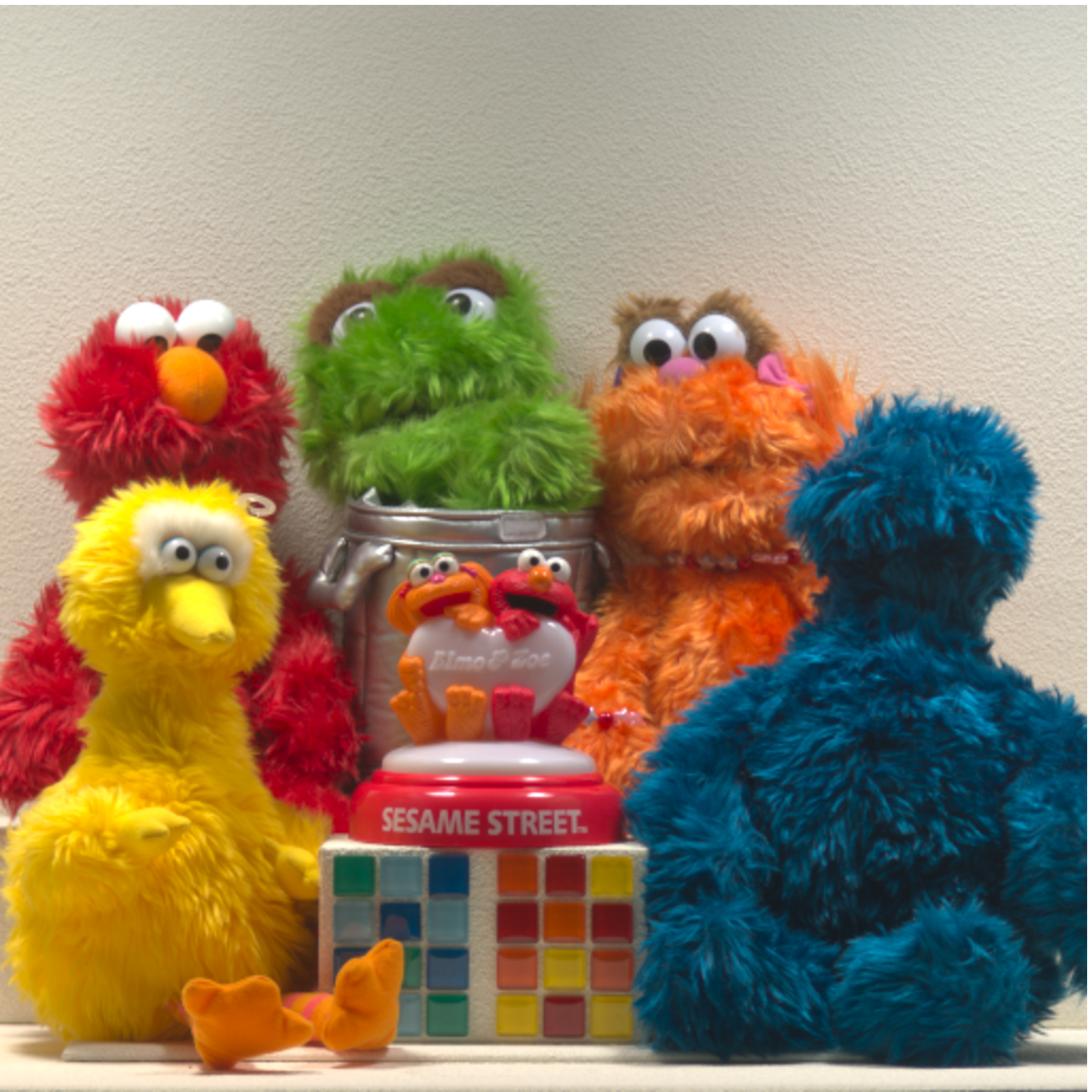}%
  \label{fig:Toys}}
 }
 \caption{Test images with $512 \times 512$ pixels, 12 bits/pixel, and 16 bands. (a) {\it Dishes} (b) {\it Scarf} (c) {\it Toys}.}
 \label{fig:TestImages}
\end{figure}

\begin{figure}[!t]
 \begin{center}
  \includegraphics[width = 0.9\linewidth]{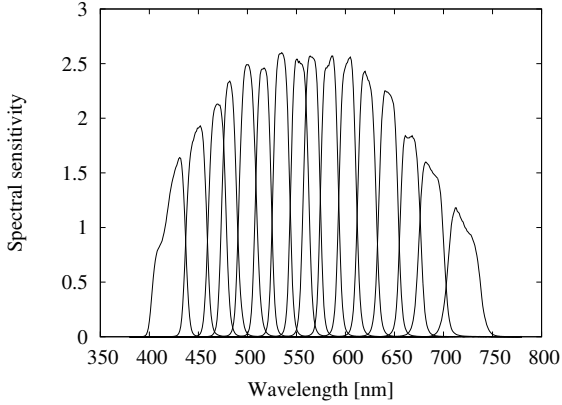}
 \end{center}
 \caption{Spectral sensitivities of color filters. The centers of the wavelengths are \{424, 448, 469, 482, 500, 517, 535, 554, 566, 584, 602, 622, 644, 666, 687, and 720\} nm. All filters have full resolution (not to mosaic) in the spatial domain.}
 \label{fig:Filter16}
\end{figure}

\begin{figure}[!t]
 \centerline{
  \subfloat[]{\includegraphics[width = 0.28\linewidth]{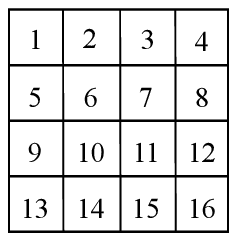}%
  \label{fig:Raster}}
  \hfil
  \subfloat[]{\includegraphics[width = 0.28\linewidth]{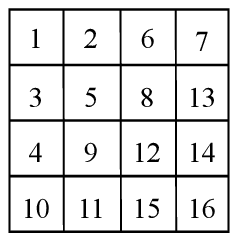}%
  \label{fig:Zigzag}}
  \hfil
  \subfloat[]{\includegraphics[width = 0.28\linewidth]{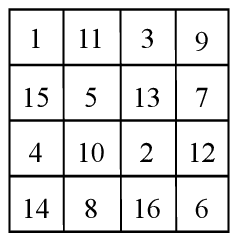}%
  \label{fig:Dither}}
 }
 \caption{Arrangement of 16-band MSFA. The number on the grid denotes a band index. A smaller index means a shorter wavelength. (a) Raster (b) Zig-zag (c) Dither.}
 \label{fig:ScanOrder}
\end{figure}

\begin{figure}[!t]
 \centerline{
  \subfloat[]{\includegraphics[width=0.87\linewidth]{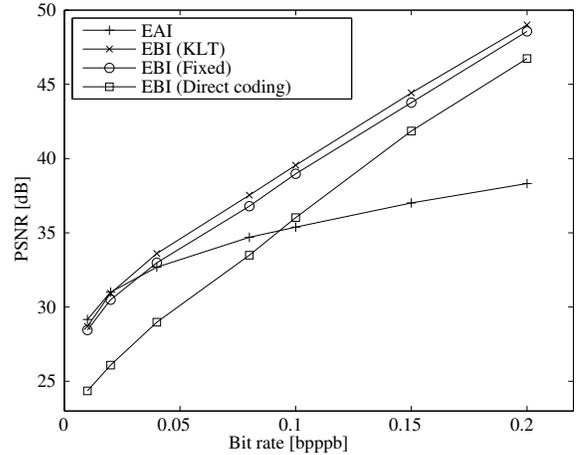}%
  \label{fig:RD_Dishes}}
 }
 \centerline{
  \subfloat[]{\includegraphics[width=0.87\linewidth]{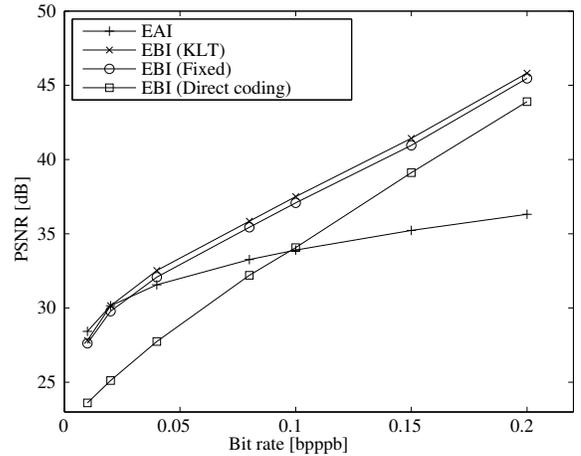}%
  \label{fig:RD_Scarf}}
 }
 \centerline{
  \subfloat[]{\includegraphics[width=0.87\linewidth]{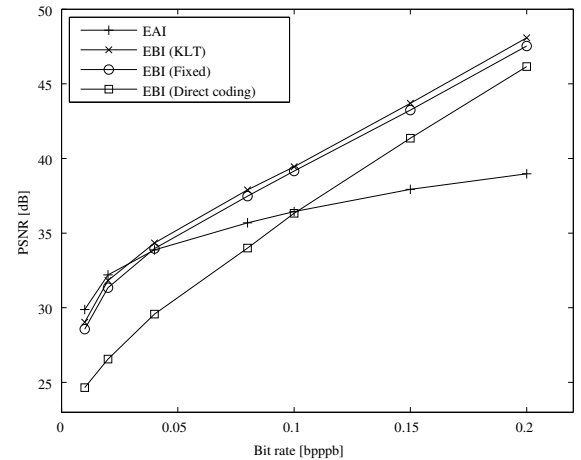}%
  \label{fig:RD_Toys}}
 }
 \caption{PSNR {\it vs}. bit rate (bit/pixel/band, bpppb) for various images. (a) {\it Dishes} (b) {\it Scarf} (c) {\it Toys}.}
 \label{fig:RD_image}
\end{figure}

In this section, we compare the compression performance of EAI and EBI by using the test images shown in \figref{fig:TestImages}.
We assign the band number to a captured image in the ascending order of wavelength in \figref{fig:Filter16} (e.g., a captured image with the 424-nm filter is band 1).
A mosaicked image is obtained by masking the captured MSI with MSFA in the simulation.
In other words, a certain pixel of the mosaicked image is generated by choosing a certain band at each pixel position from the captured MSI.
The MSFA patterns used in the experiment are shown in \figref{fig:ScanOrder}, and the obtained mosaicked image is demosaicked by J.~Brauers' method \cite{ref:JBrauers2006}.

A full-resolution EAI image and a transformed EBI image are encoded by using Jasper \cite{ref:Jasper} with 9/7 real wavelet transform.
We use KLT as a spectral transform for EAI.
The peak signal-to-noise ratio (PSNR) of a decoded MSI in EAI and EBI is calculated with a demosaicked image.
For the fixed transform matrix of EBI, $\rho_{_{f}}$ and $\rho_{_{d}}$ are set to 0.995 and 0.95, respectively.

The experimental results are affected by various factors, such as the test images shown in \figref{fig:TestImages}, MSFA patterns shown in \figref{fig:ScanOrder}, and the number of bands.
Therefore, first, a basic comparison between EAI and EBI is shown in \secref{ssec:PSNR_image} with a dither MSFA pattern (\figref{fig:ScanOrder}c) of 16 bands.
Then, we examine the compression performance by changing the simulation conditions in order to confirm the robustness of the proposed method.
The performance comparison of three MSFA patterns is shown in \secref{ssec:PSNR_pattern}.
\secref{ssec:PSNR_num} shows the performance difference when changing the number of bands.
Finally, we compare the two different calculation methods for PSNR in \secref{ssec:PSNR_OPSNR}.

\begin{figure*}[!t]
 \centerline{
  \subfloat[]{\includegraphics[width=0.3\linewidth]{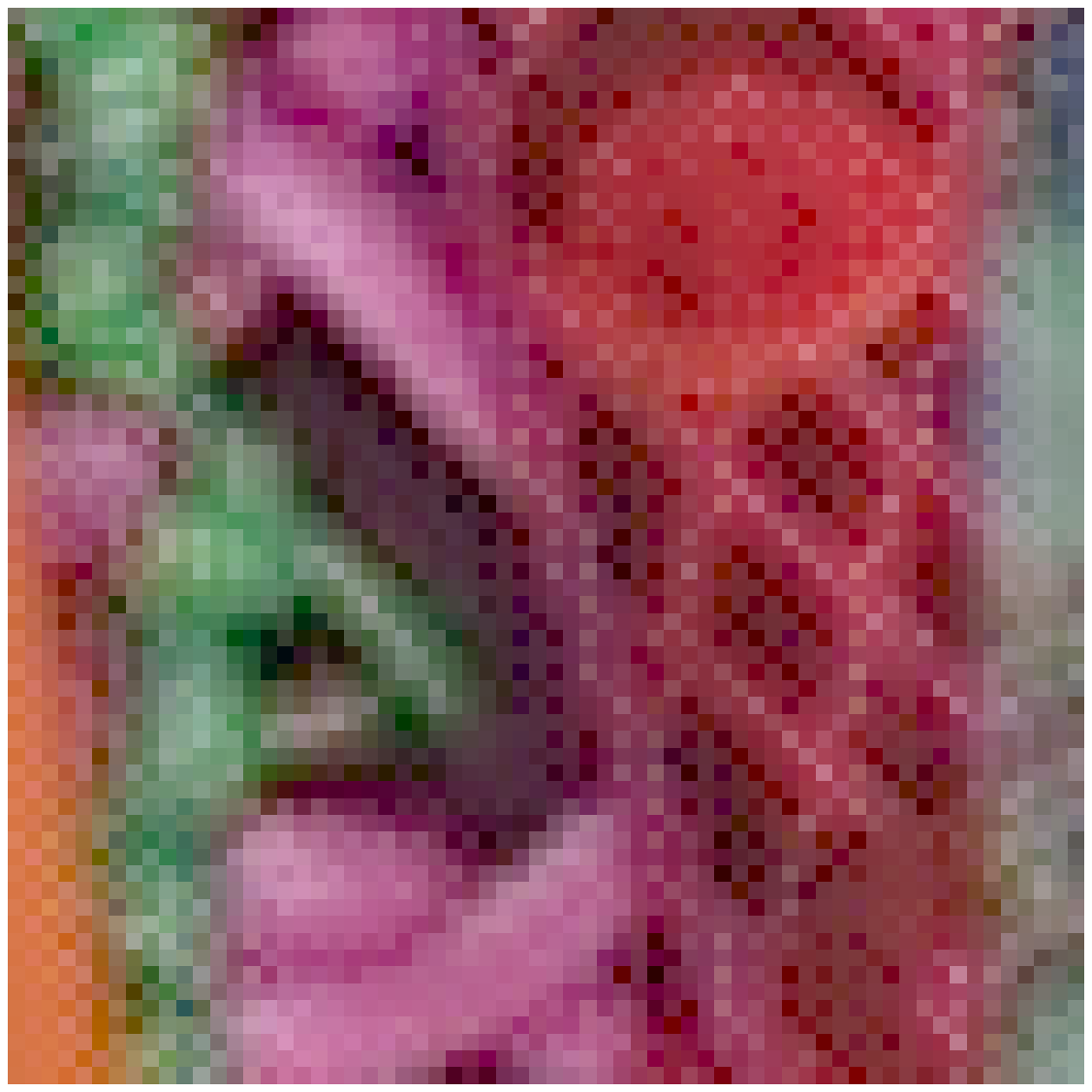}%
  \label{fig:dec_image0.100_01}}
  \hfil
  \subfloat[]{\includegraphics[width=0.3\linewidth]{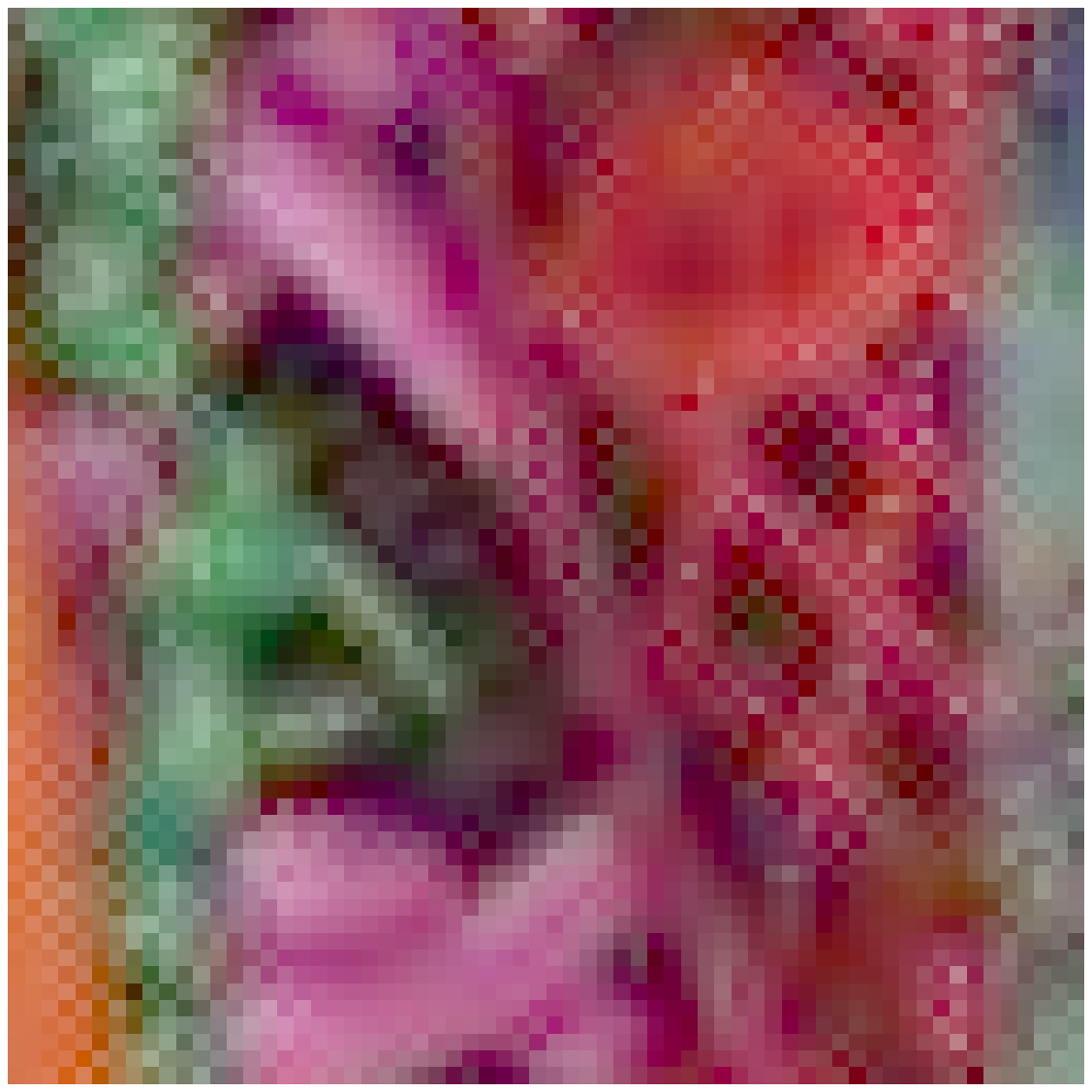}%
  \label{fig:dec_image0.100_02}}
 }
 \centerline{
  \subfloat[]{\includegraphics[width=0.3\linewidth]{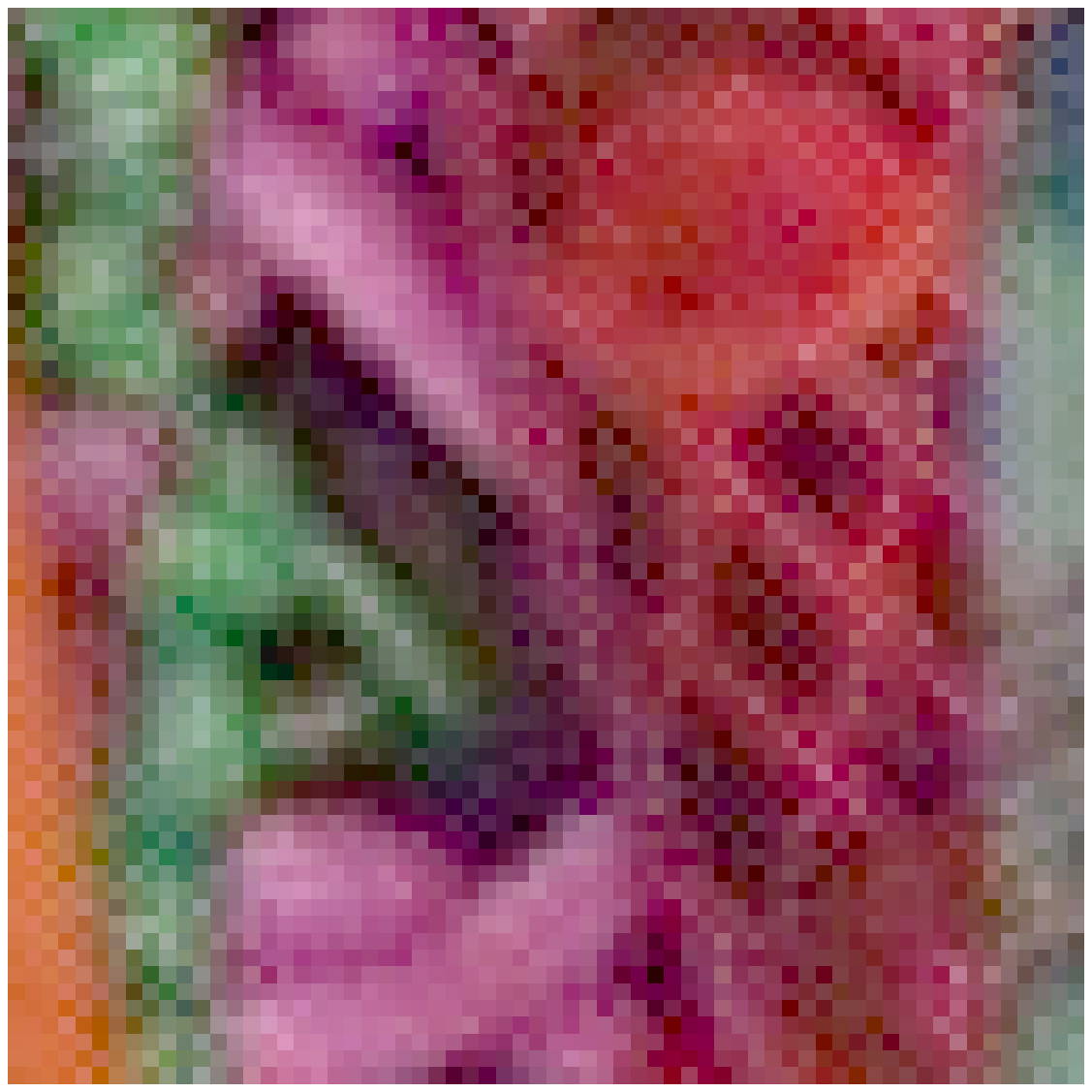}%
  \label{fig:dec_image0.100_03}}
  \hfil
  \subfloat[]{\includegraphics[width=0.3\linewidth]{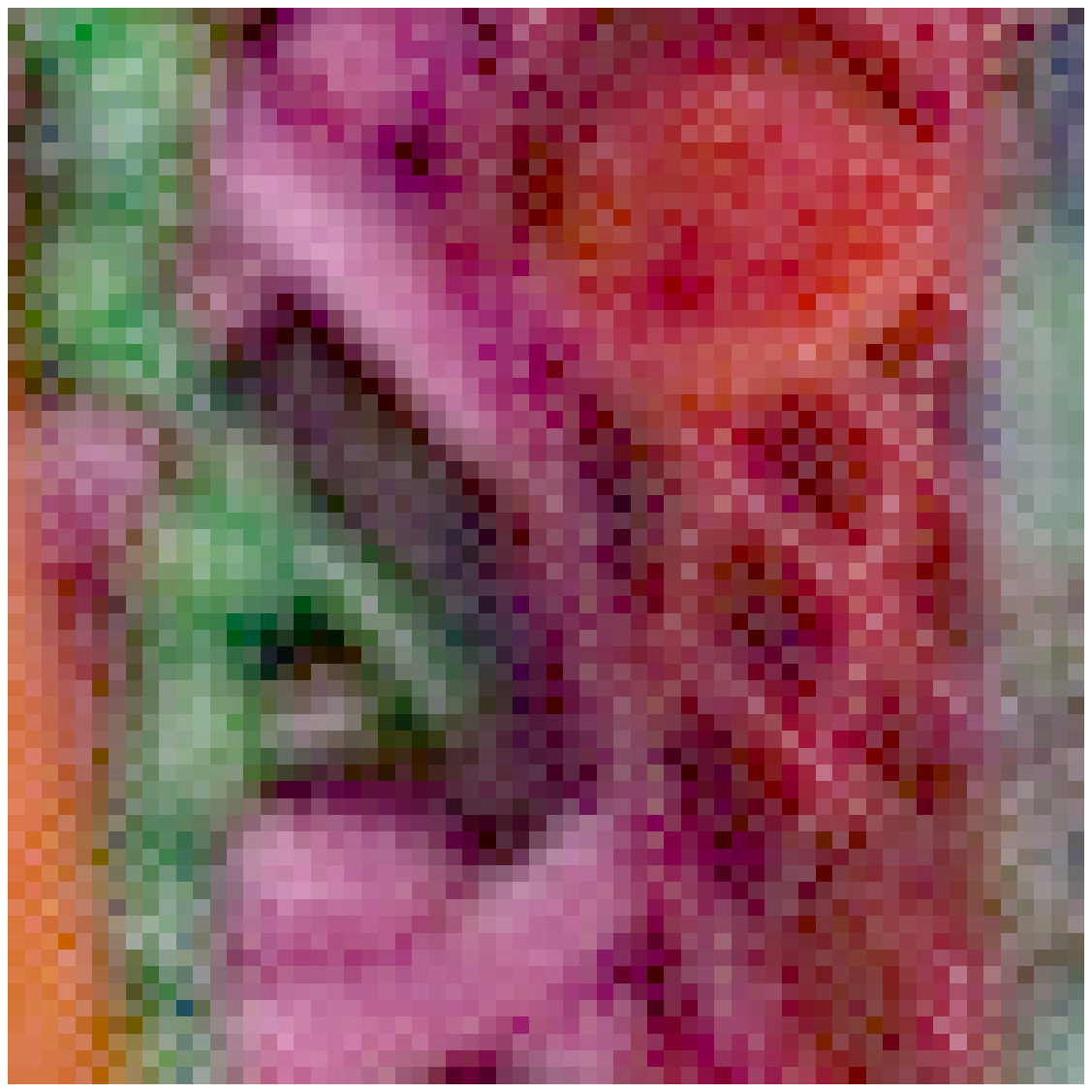}%
  \label{fig:dec_image0.100_04}}
 }
 \caption{Comparison of {\it Scarf} images at 0.1 bpppb. (a) Demosaicked without compression (b) EAI (c) EBI (KLT) (d) EBI (fixed).}
 \label{fig:dec_image0.100}
\end{figure*}

\subsection{Comparison of three test images} \label{ssec:PSNR_image}

\figref{fig:RD_image} shows the comparison of PSNR against the bit rate between EAI, EBI (KLT), EBI (Fixed), and EBI (Direct coding) for three test images.
EBI (KLT) means that the KLT was chosen using all samples of the image, and EBI (Fixed) means that the KLT was chosen using the fixed transform matrix described in \eqref{eq:FixedMatrixTransform}.
First, we note that EBI (KLT) outperforms EAI at almost all bit rates in \figref{fig:RD_image}.
This graph shows that EBI can contribute greatly to the data reduction of MSI.
Although EAI is better than EBI (KLT) for some lower bit rates, the difference is not large in all cases.
This trend is similar to what was observed in \cite{ref:NXLian2006,ref:SYLee2009} for Bayer image compression.
EAI and EBI differ in both the number of wavelet coefficients to be encoded and the distribution of coefficient energy over different frequencies. Thus, at high bit rates, when most wavelet coefficients are being refined (most are already significant), EAI is at a disadvantage because it has to encode $N$ times more coefficients than EBI.
Conversely, at low bit rates, the behavior is dominated by the distribution of the few wavelet coefficients that have already been declared significant.
At low rates, EAI performs better because it has lower frequencies than EBI in both the spectral and the spatial domain, and thus, the same amount of signal energy can be compacted into a smaller number of coefficients in EAI than in EBI.

Comparing EBI (KLT) and EBI (Fixed) in \figref{fig:RD_image},
we find that EBI (Fixed) is comparable to EBI (KLT) and outperforms EBI (Direct coding).
This implies that we can select a fixed transform matrix for a multispectral camera in the manufacturing phase with minimal loss in PSNR.
\figref{fig:RD_Dishes} shows a slightly larger difference between EBI (KLT) and EBI (Fixed) than \figref{fig:RD_Scarf} and \figref{fig:RD_Toys} because the PSNR difference between KLT and the fixed transform may depend on the correlation coefficients, contents, spectral distribution, and any other image structure.

EBI (Direct coding) is clearly not competitive with the other EBI methods in terms of the rate-distortion performance.
The cross-over point of PSNR between EBI (Direct coding) and EAI can be seen around 0.1 bpppb in all images,
 but EBI (Direct coding) has few advantages except for the calculation cost.
Therefore, the performance of EBI (Direct coding) is not considered in the following experiments.

\figref{fig:dec_image0.100} shows the decoded and demosaicked {\it Scarf} images at 0.1 bpppb.
\figref{fig:dec_image0.100_01} shows the demosaicked image without compression, which is referred to as the original image.
\figref{fig:dec_image0.100_02} shows the image decoded by EAI; some blurring effect can be seen here because of both demosaicking and compression.
\figsref{fig:dec_image0.100_03}{fig:dec_image0.100_04} obtained by EBI have similar qualities and show a superior image than \figref{fig:dec_image0.100_02}.

\subsection{Comparison of MSFA patterns} \label{ssec:PSNR_pattern}

\begin{figure}[!t]
 \begin{center}
  \includegraphics[width = 0.9\linewidth]{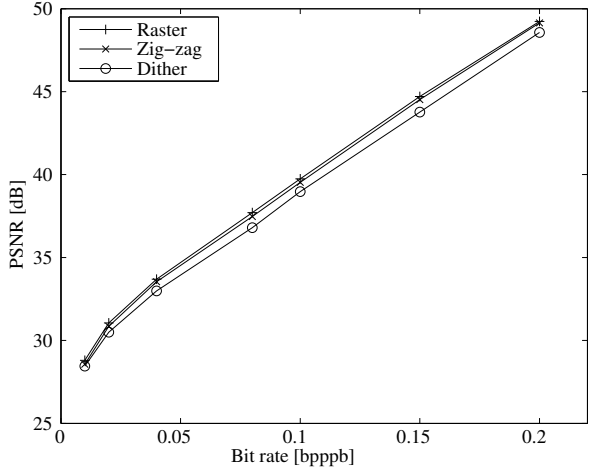}
 \end{center}
 \vspace{-3mm}
 \caption{PSNR {\it vs}. bit rate for various MSFA patterns in {\it Dishes}.}
 \label{fig:RD_pattern}
\end{figure}

\begin{table}[!t]
 \begin{center}
  \caption{Comparison of coding gain in a fixed transform matrix of EBI.}
  \label{tab:CodingGain}
\begin{tabular}{c||c|c|c} \hline
MSFA pattern & Raster & Zig-zag & Dither \\ \hline \hline
Coding gain [dB] & 9.441 & 9.379 & 8.709 \\ \hline
 \end{tabular}
 \end{center}
\end{table}

\begin{figure}[!t]
 \centerline{
  \subfloat[]{\includegraphics[width=0.9\linewidth]{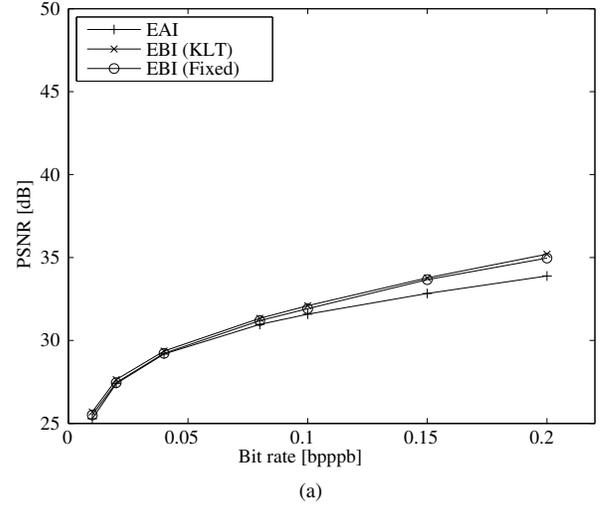}%
  \label{fig:RD_3band}}
  }
 \centerline{
  \subfloat[]{\includegraphics[width=0.9\linewidth]{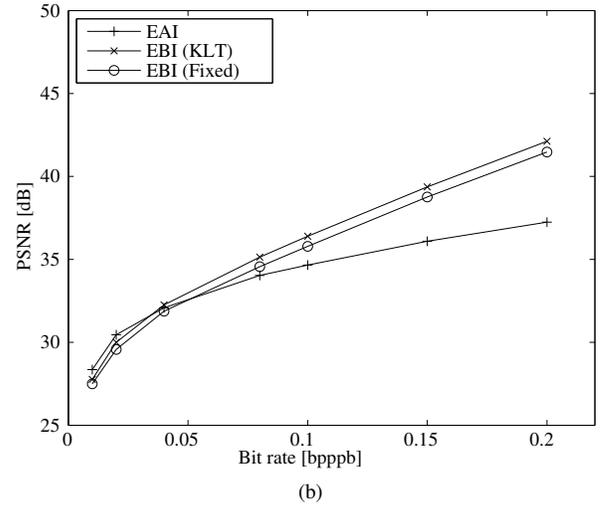}%
  \label{fig:RD_9band}}
 }
 \centerline{
  \subfloat[]{\includegraphics[width=0.9\linewidth]{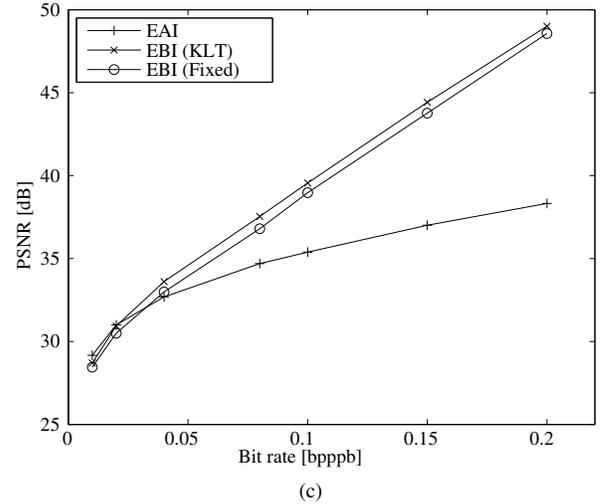}%
  \label{fig:RD_16band}}
 }
 \caption{PSNR {\it vs}. bit rate in {\it Dishes}. (a) 3 bands (b) 9 bands (c) 16 bands.}
 \label{fig:RD_NumOfBands}
\end{figure}

A comparison of the different MSFA patterns is shown in \figref{fig:RD_pattern}, which shows that the dither pattern has lower performance than other MSFAs.
This is because {\it structure conversion} is based on the order of the center wavelength.
To confirm this, coding gains of the fixed transform matrix on each MSFA are shown in \tabref{tab:CodingGain}.
The coding gain $G$ is calculated using eigenvalues corresponding to the SVD of $\Vec{R}_{fd}$ as follows:
\begin{eqnarray}
 G = 10 \log_{10}\frac{\displaystyle\frac{1}{N}\displaystyle\sum_{i=1}^{N}\lambda_{i}}{\left(\displaystyle\prod_{i=1}^{N}\lambda_{i}\right)^{\frac{1}{N}}} = 10 \log_{10}\frac{1}{\textrm{det}(\Vec{R}_{fd})^{\frac{1}{N}}},
\end{eqnarray}
where $\lambda_{i}$ denotes the eigenvalue corresponding to the SVD of $\Vec{R}_{fd}$ and $\textrm{det}(\Vec{R}_{fd})$ represents the determinant of $\Vec{R}_{fd}$.
The coding gain of the Raster pattern shows the highest value because the neighboring filters on the spectrum are well-connected at the spatial position.
Although the dither pattern has advantages in terms of the demosaicking quality \cite{ref:KShinoda2013}, its coding gain is the lowest.
From the perspective of crosstalk, neighboring filters on the spectrum (e.g., ${\rm S}^{(n)}$ and ${\rm S}^{(n+1)}$) should be arranged distant from each other in order to improve the demosaicked quality.
However, from the perspective of EBI compression, each of them should be arranged close in order to improve the coding gain.
The relation between reconstruction and compression performances is a trade-off.
If there is a high degree of freedom in selecting the arrangement of MSFA within the manufacturing process, then filters that are neighbors on the spectrum should be arranged close to each other in order to improve the compression performance.

\subsection{Comparison using various numbers of bands} \label{ssec:PSNR_num}

\figref{fig:RD_NumOfBands} shows a comparison using various numbers of bands.
The size of the MSFA block is assumed to be one of the following three types: $4 \times 4$ pixels for 16 bands; $3 \times 3$ pixels for 9 bands \{424, 469, 500, 535, 566, 584, 622, 666, and 720 nm\}, where filters are chosen at regular intervals from \figref{fig:Filter16}; and $2 \times 2$ Bayer CFA for 3 bands.
These three images are generated from the same 16-band image, but each of the images has different bands.

From \figref{fig:RD_NumOfBands}, it is clear that EBI has an advantage over EAI in terms of higher bit rates irrespective of the number of bands.
With an increase in the number of bands, the number of encoded pixels increases in the case of EAI, whereas the number remains unchanged in the case of EBI.
Therefore, applying EBI to a multispectral imaging system has an advantage as the number of bands is increased.

\subsection{Comparison using different distortion metrics} \label{ssec:PSNR_OPSNR}

\begin{figure}[!t]
 \centerline{
  \subfloat[]{\includegraphics[width=1\linewidth]{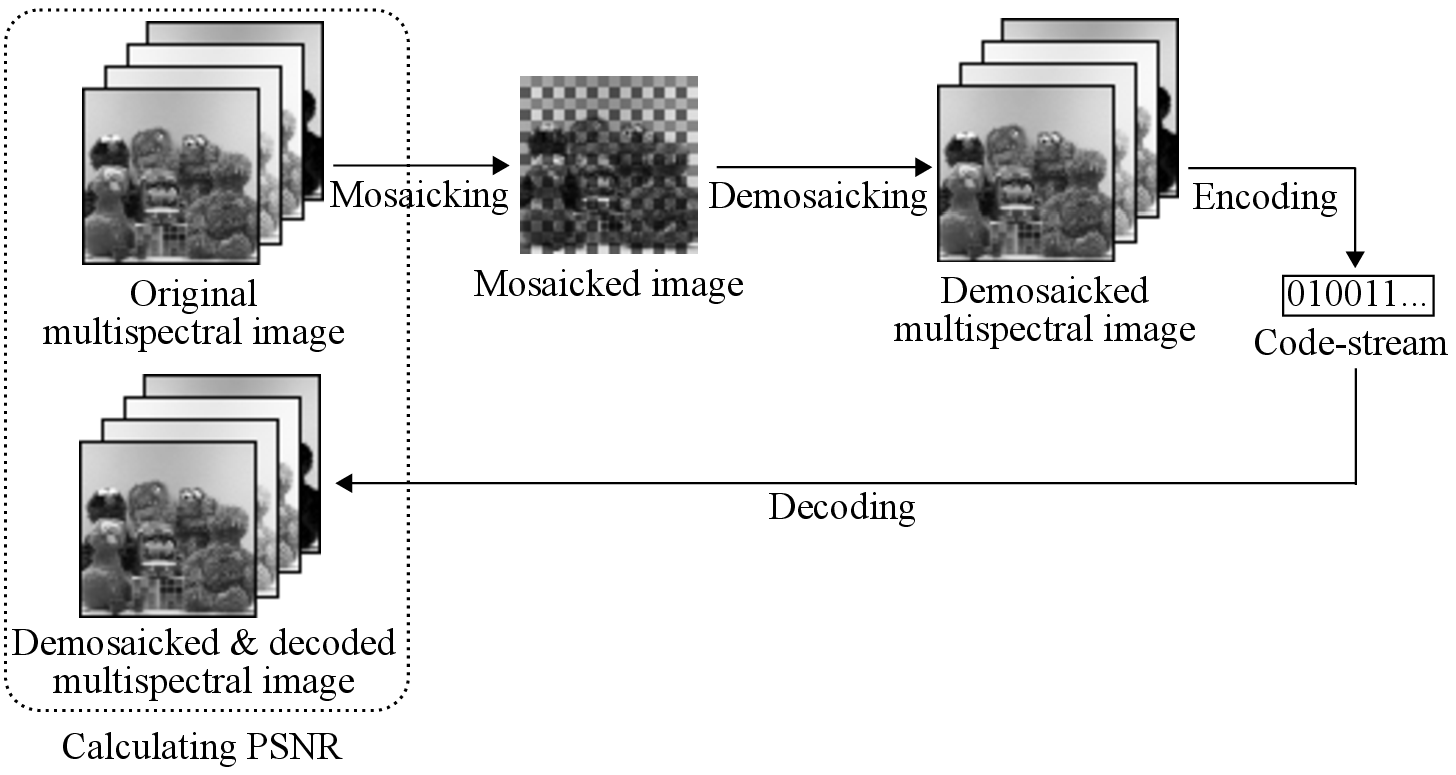}%
  \label{fig:OPSNR_EAI}}
 }
 \centerline{
  \subfloat[]{\includegraphics[width=1\linewidth]{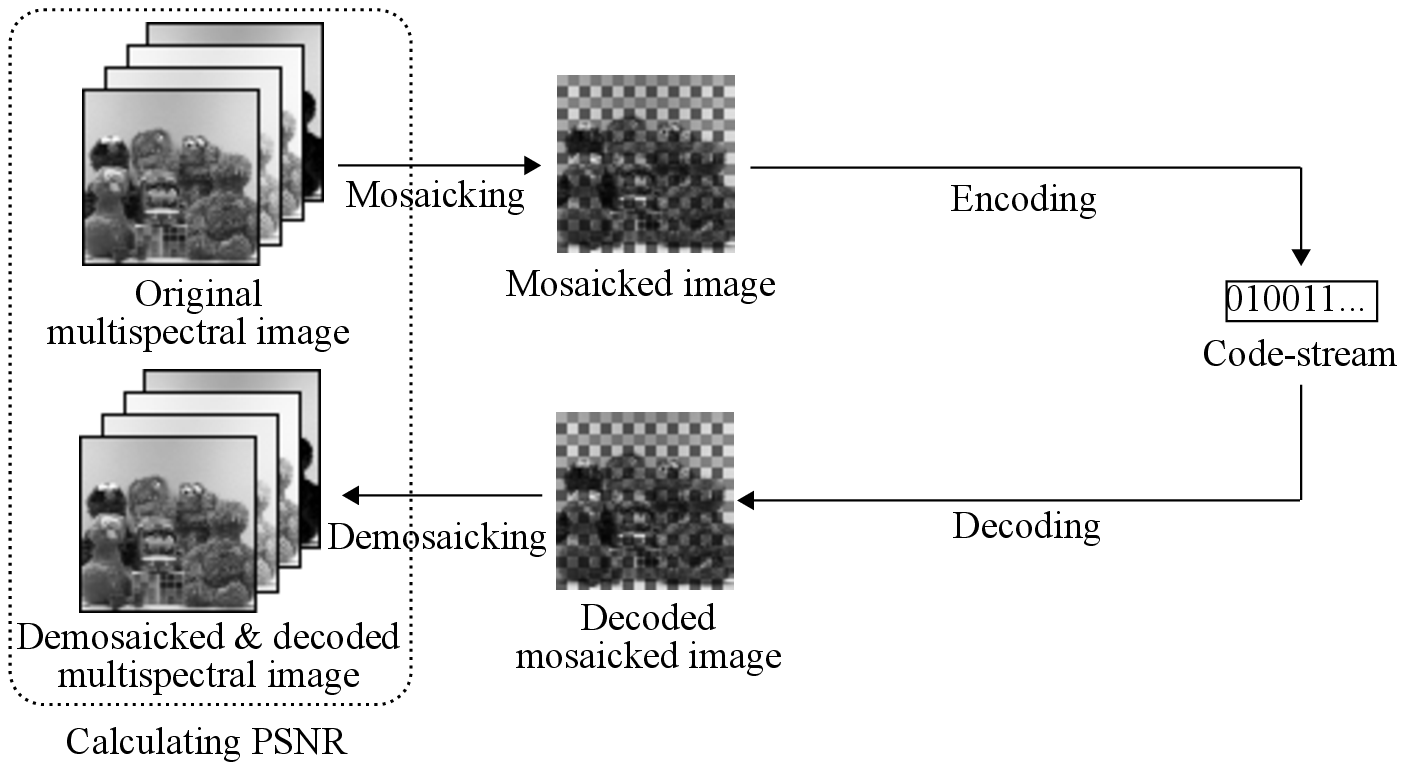}%
  \label{fig:OPSNR_EBI}}
 }
 \caption{OPSNR calculation. OPSNR is calculated between ``original'' and ``demosaicked \& decoded multispectral" images. (a) In the case of EAI (b) In the case of EBI.}
 \label{fig:OPSNR}
\end{figure}

\begin{figure}[!t]
 \centerline{
  \subfloat[]{\includegraphics[width=1\linewidth]{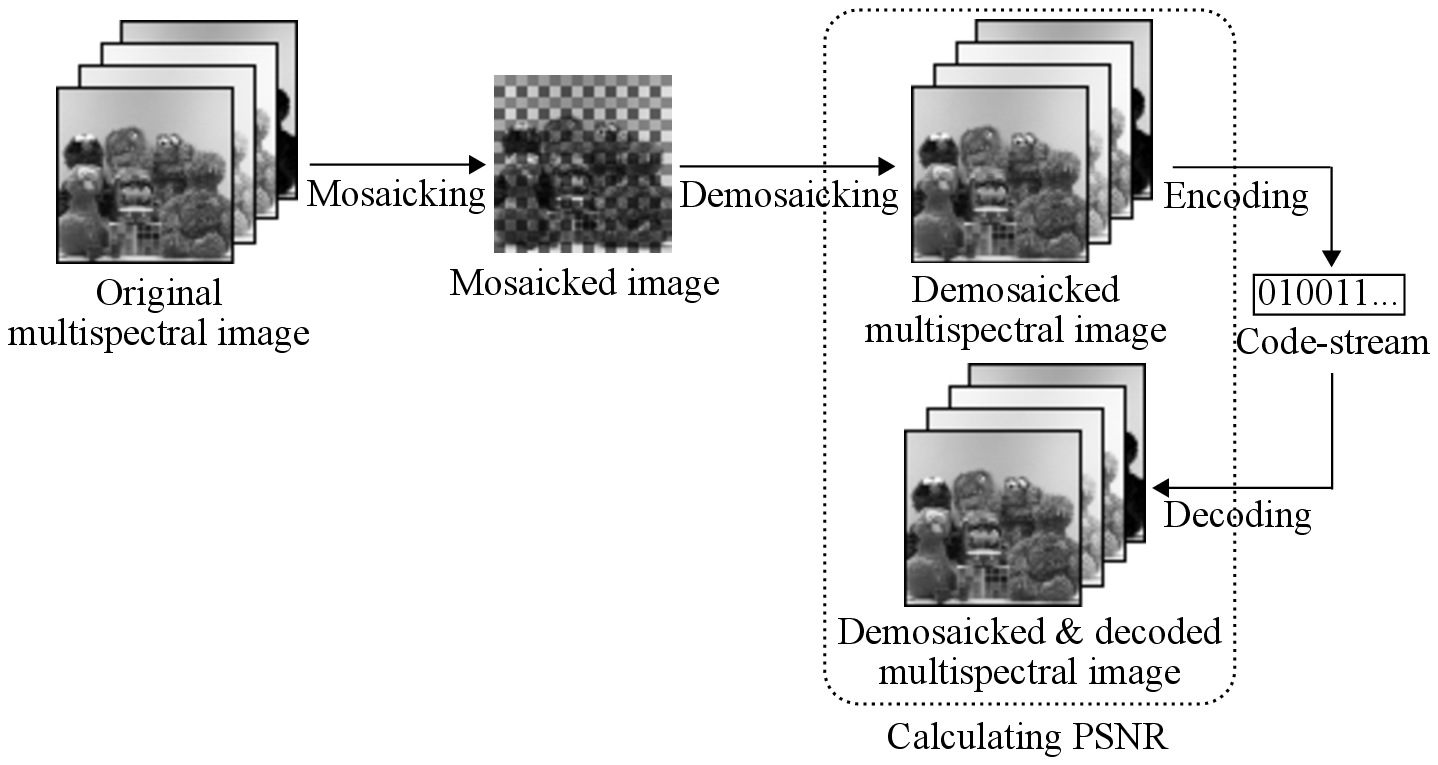}%
  \label{fig:DPSNR_EAI}}
 }
 \centerline{
  \subfloat[]{\includegraphics[width=1\linewidth]{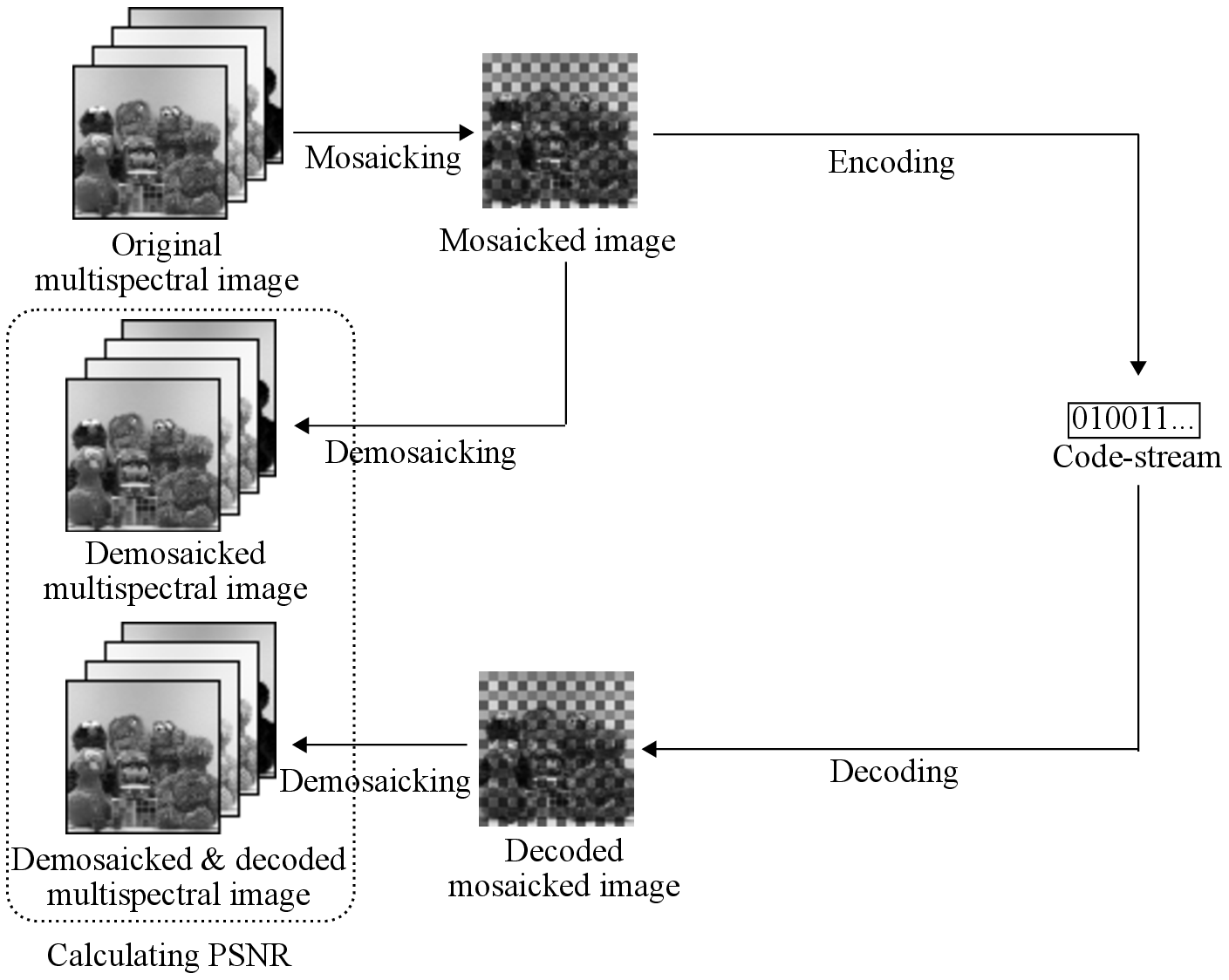}%
  \label{fig:DPSNR_EBI}}
 }
 \caption{DPSNR calculation. DPSNR is calculated between ``demosaicked'' and ``demosaicked \& decoded multispectral" images. (a) In the case of EAI (b) In the case of EBI.}
 \label{fig:DPSNR}
\end{figure}

We use two PSNR metrics as shown in \figsref{fig:OPSNR}{fig:DPSNR}.
First, we use as the reference, the original full-resolution image, which is obtained before MSFA (\figref{fig:OPSNR}).
We call this comparison ``original-PSNR (OPSNR),'' which means ``PSNR calculated with an original full-resolution image.''
OPSNR can be calculated in a simulated system as shown in \figref{fig:OPSNR} but cannot be calculated in a real system because the original full-resolution image is not available.
As an alternative, we use as the reference the demosaicked image obtained from the original pixels (i.e., non-compressed) (\figref{fig:DPSNR}).
We call this comparison ``demosaicked-PSNR (DPSNR)'', i.e., ``PSNR calculated with a demosaicked full-resolution image.''
Although the true image of DPSNR includes distortion due to down-sampling and demosaicking, the demosaicked image can be obtained by using either a simulated system or a real multispectral single-sensor camera.
In the previous studies, \cite{ref:SYLee2001,ref:BParrein2004,ref:XXie2005} used DPSNR, whereas \cite{ref:NXLian2006} mainly used OPSNR.
In this study, we mainly used DPSNR, but the validation of OPSNR is also considered important in the multispectral case.

\begin{figure}[!t]
 \centerline{
  \subfloat[]{\includegraphics[width=0.9\linewidth]{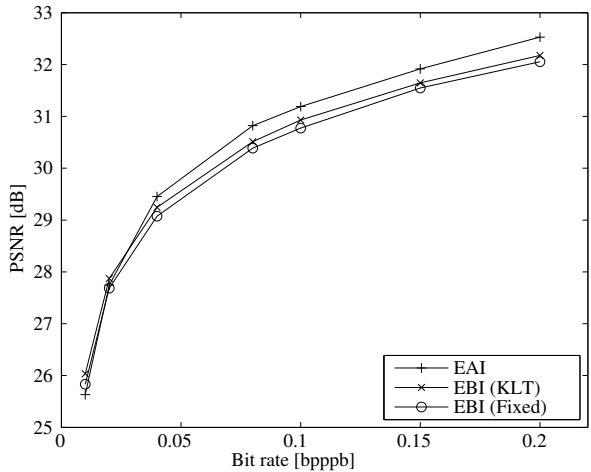}%
  \label{fig:RD_OPSNR_3band}}
}
 \centerline{
  \subfloat[]{\includegraphics[width=0.9\linewidth]{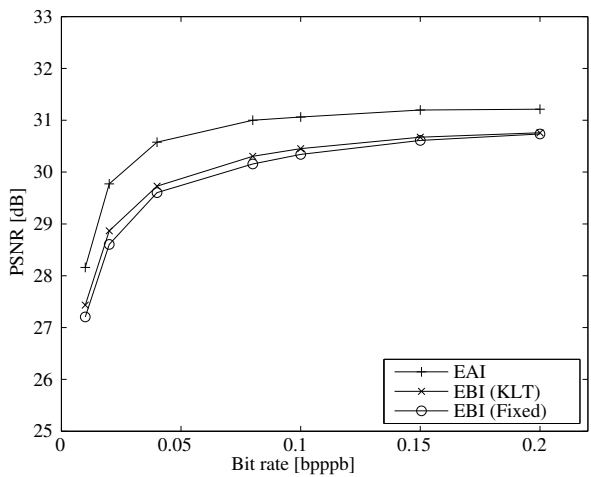}%
  \label{fig:RD_OPSNR_9band}}
 }
 \centerline{
  \subfloat[]{\includegraphics[width=0.9\linewidth]{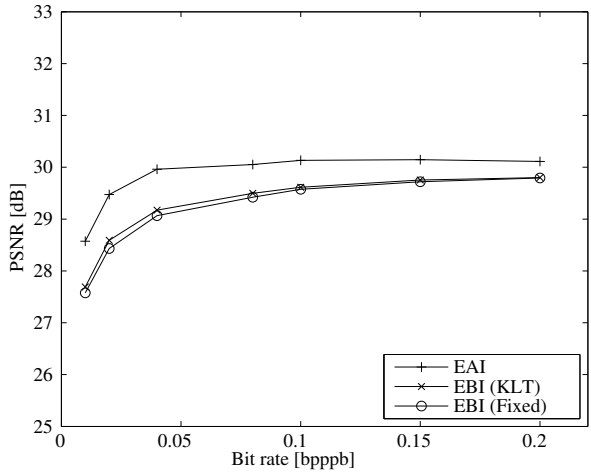}%
  \label{fig:RD_OPSNR_16band}}
 }
 \caption{OPSNR {\it vs}. bit rate in {\it Dishes}. (a) 3 bands (b) 9 bands (c) 16 bands.}
 \label{fig:RD_OPSNR}
\end{figure}


\begin{table}[!t]
 \begin{center}
  \caption{Comparison of PSNR between original and demosaicked images without compression.}
  \label{tab:OPSNR_ls}
\begin{tabular}{c||c|c|c} \hline
Number of bands & 3 & 9 & 16 \\ \hline
PSNR [dB] & 57.876 & 30.956 & 29.830 \\ \hline
 \end{tabular}
 \end{center}
\end{table}

\figref{fig:RD_OPSNR} shows OPSNR for different numbers of bands.
The PSNR does not become arbitrarily large as the bit rate increases because it includes the error between the original and the demosaicked image.
The demosaicked image quality without compression is shown in \tabref{tab:OPSNR_ls}.
Clearly, in \figref{fig:RD_OPSNR}, we see that the corresponding PSNRs in \tabref{tab:OPSNR_ls} are reached asymptotically.

From \figref{fig:RD_OPSNR}, we observe that the PSNR of EBI is smaller than that of EAI, and it is almost the same result as that mentioned by Lian {\it et al.} \cite{ref:NXLian2006}.
It is difficult to identify the cause of the performance difference because OPSNR is related to not only compression and interpolation but also MSFA.
Lian {\it et al.} \cite{ref:NXLian2006} studied the theoretical EBI model and compared EAI and EBI using a simulation.
Their results show that EBI outperforms EAI at higher bit rates but underperforms at lower bit rates.

Note that we cannot conclude whether OPSNR or DPSNR is a better evaluation when comparing EAI and EBI.
OPSNR can be calculated in a computational simulation, but OPSNR cannot be calculated in a real MSFA-based system because the original full-resolution image does not exist.
On the other hand, DPSNR is a method to measure quality based on a demosaicked image in a real MSFA-based system.

Further, in a comparison of these two measurements, OPSNR includes the effect of mosaicking process by MSFA, whereas DPSNR does not include it.
OPSNR cannot reach infinity because of the error of the mosaicking-to-demosaicking process.
This limitation may affect the EBI performance at higher bit rates.
Although the results of OPSNR can be improved by modifying MSFA and demosaicking, this work is outside the scope of this paper.
This paper focuses on compression methods for mosaicked MSI; hence, DPSNR has been used mainly as a metric for the performance of EAI and EBI.
Since the OPSNR performance of the proposed method cannot be improved unless we consider the relation between MSFA, demosaicking, and compression, the OPSNR evaluation under other experimental conditions is omitted.

\section{Discussion} \label{sec:Discussion}

In this section, we discuss the first-order correlation coefficients ($\rho_{_{f}}$ and $\rho_{_{d}}$) of the proposed fixed transform.
In the practical use of multispectral cameras, various applications are assumed to be used in the future.
If original images can be obtained before producing MSFA, $\rho_{_{f}}$ and $\rho_{_{d}}$ can be adjusted to appropriate values.
As a more effective approach, instead of these correlation coefficients, an offline KLT matrix is generated using the original images.
However, it is difficult to obtain original images before producing MSFA in many cases.
Therefore, specifying $\rho_{_{f}}$ and $\rho_{_{d}}$ without original images is an important challenge.

Although we set $\rho_{_{f}}$ and $\rho_{_{d}}$ to a value that is referred from \cite{ref:WKPratt1976,ref:RJClarke1981,ref:Spr2009,ref:Murakami2009} in \secref{ssec:Fixed}, a critical inconsistency was not observed in our verification.
Therefore, the proposed fixed transform has the potential to work efficiently even in the case of using a slightly different first-order correlation coefficient.
We show the validation of the fixed transform by using a different MSI and different first-order correlation coefficients in this section.

\begin{figure}[!t]
 \begin{center}
  \includegraphics[width=0.9\linewidth ,scale=0.3]{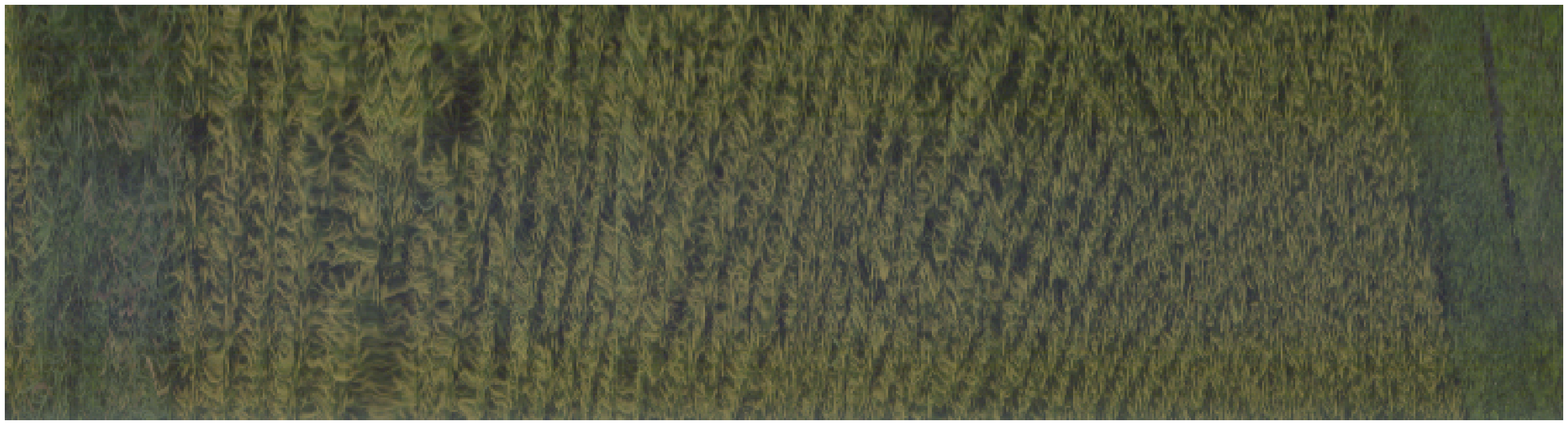}
 \end{center}
 \caption{{\it Rice field} with $1500 \times 400$ pixels and 121 bands.}
 \label{fig:VegetationImage}
\end{figure}

\begin{figure}[!t]
 \begin{center}
  \includegraphics[width = 0.85\linewidth]{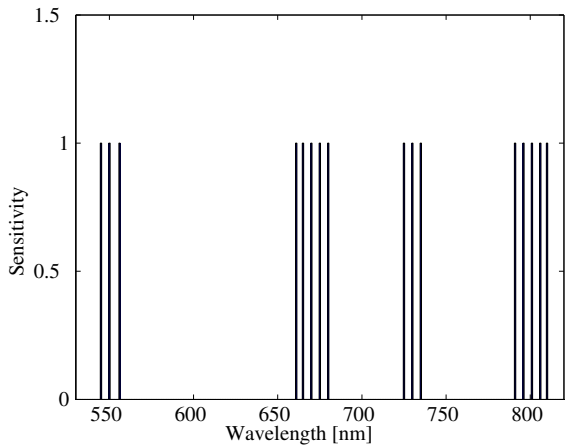}
 \end{center}
 \caption{Assumed multispectral filter specification in {\it Rice field}. The centers of the wavelength are
  \{545, 550, 556, 661, 665, 670, 675, 680, 725, 730, 735, 791, 796, 801, 805, and 810\} nm.
 All filters have full resolution in the spatial domain.}
 \label{fig:Filter16_vegetation}
\end{figure}


\begin{table}[!t]
 \begin{center}
  \caption{First-order correlation coefficients between the test images.}
  \label{tab:RealCorrelationCoeff}
\begin{tabular}{c||c|c|c|c} \hline
Test image & {\it Dishes} & {\it Scarf}  & {\it Toys} & {\it Rice field} \\ \hline\hline
$\rho_{_{f}}$ & 0.9994 & 0.9978 & 0.9970  & 0.9991 \\ \hline
$\rho_{_{d}}$ & 0.9108 & 0.9601 & 0.9671 & 0.9068 \\ \hline
  \end{tabular}
 \end{center}
\end{table}

\begin{figure}[!t]
 \begin{center}
  \includegraphics[width=0.9\linewidth]{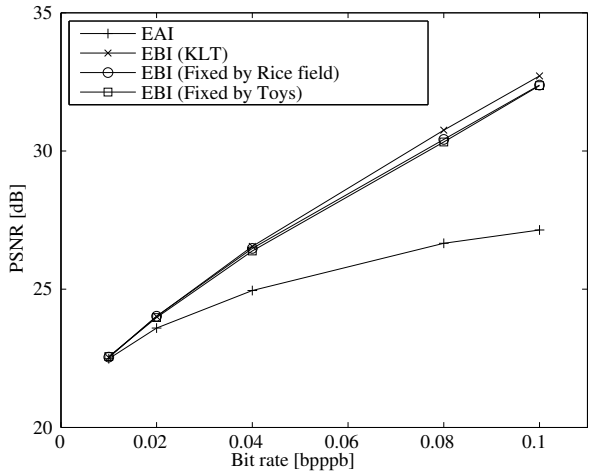}
 \end{center}
 \caption{PSNR {\it vs}. bit rate in {\it Rice field}.}
 \label{fig:RD_Vegetation}
\end{figure}

We use an additional test image, which is shown in \figref{fig:VegetationImage}({\it Rice field}). This image is captured by a hyperspectral camera equipped with 121 filters.
The wavelength range of the filters is 400 to 1000 nm, and the wavelength resolution is 5 nm.
We choose 16 bands from {\it Rice field} for the following experiments, as shown in \figref{fig:Filter16_vegetation}.
These wavelengths are selected based on the vegetation analysis \cite{ref:Tucker1979}.
The camera-to-subject distance is a few dozen meters in \figref{fig:VegetationImage}.
\tabref{tab:RealCorrelationCoeff} shows the real first-order correlation coefficients.
$\rho_d$ of an image is calculated from full-resolution MSIs as follows:
\begin{eqnarray}
 \rho_{d} = \frac{1}{N}\sum^{N}_{n=1}\left\{ \frac{\displaystyle\sum^{M}_{i=2}(x_{_{n,i}}-\overline{x_n})(x_{_{n,i-1}}-\overline{x_{_{n}}})}{\displaystyle\sum^{M}_{i=1}(x_{_{n,i}}-\overline{x_{_{n}}})^{2}} \right\}, \label{eq:real_rhod}
\end{eqnarray}
where $x_{_{n,i}}$ denotes a pixel value of band $n$ and spatial index $i$, which is converted from a 2-D image to a 1-D signal, $\overline{x}$ represents an average value of band $n$, and $M$ indicates the number of pixels per band.
\eqref{eq:real_rhod} corresponds to the average of the first-order auto-correlation coefficients in the spatial domain.
$\rho_{_{f}}$ is calculated as
\begin{eqnarray}
 \rho_{_{b(n,n-1)}} &\hspace{-2mm}=&\hspace{-2mm} \frac{\displaystyle\sum^{M}_{i=1}(x_{_{n,i}}-\overline{x_{_{n}}})(x_{_{n-1,i}}-\overline{x_{_{n-1}}})}{\sqrt{\displaystyle\sum^{M}_{i=1}(x_{_{n,i}}-\overline{x_{_{n}}})^{2}}\sqrt{\displaystyle\sum^{M}_{i=1}(x_{_{n-1,i}}-\overline{x_{_{n-1}}})^{2}}} \label{eq:real_rhob} \\
 \rho_{_{f}} &\hspace{-2mm}=&\hspace{-2mm} \frac{1}{N-1}\sum^{N}_{n=2} \rho_{_{b(n,n-1)}}^{^{\frac{1}{f_{n,n-1}}}}, \label{eq:real_rhof}
\end{eqnarray}
where \eqref{eq:real_rhob} corresponds to the first-order correlation coefficient between band indices in the spectral domain (without considering the optical wavelength), and \eqref{eq:real_rhof} corresponds to the average of the first-order correlation coefficients per nanometer in the spectral domain.

\figref{fig:RD_Vegetation} shows the rate-distortion performance in the vegetation image.
Here, EBI (fixed for {\it Rice field}) means that the first-order correlation coefficient of the fixed transform matrix is set to the real correlation coefficients of {\it Rice field}, and the first-order correlation coefficient of the EBI (fixed for {\it Toys}) is set to that of {\it Toys}.
Note that EBI (fixed for{\it Toys}) uses $\rho_{_{f}}$ and $\rho_{_{d}}$ calculated from {\it Toys}, but $f$ and $d$ in \eqref{eq:Markov} and \eqref{eq:Markov2} are calculated on the basis of the wavelength interval of \figref{fig:Filter16_vegetation}.
From \figref{fig:RD_Vegetation}, we find that EBI outperforms EAI and that EBI (fixed for {\it Rice field}) has almost the same PSNR as EBI (KLT).
Moreover, EBI (fixed for {\it Toys}) has almost the same PSNR as EBI (fixed for {\it Rice field}) even though different $\rho_{_{f}}$ and $\rho_{_{d}}$ values are used.
In the calculation of \eqref{eq:Markov} and \eqref{eq:Markov2}, the proposed method considers not only $\rho_{_{f}}$ and $\rho_{_{d}}$ but also the wavelength intervals and the spatial patterns of MSFA.
The performance of the proposed method can be improved by adjusting $\rho_{_{f}}$ and $\rho_{_{d}}$ but does not degrade significantly even if $\rho_{_{f}}$ and $\rho_{_{d}}$ are not adjusted.
The advantage of the proposed method is to be able to determine the correlation coefficients $\rho_{_{f}}^{f_{m, n}}$ and $\rho_{_{d}}^{d_{m, n}}$ by using $f_{m, n}$ and $d_{m, n}$, which can be obtained from an MSFA.
If an MSFA is given, the proposed method can generate an effective spectral transform without capturing object.

\section{Conclusion} \label{sec:Conclusion}
 We have presented a new compression method for mosaicked images obtained using a one-shot multispectral camera.
Encoding a mosaicked image before interpolation shows higher PSNR than encoding a full-resolution image after interpolation at almost all bit rates. The proposed method using a transform matrix derived from MSFA specifications achieves comparable PSNR with KLT; thus, it is possible for the coding parameters to be fixed without image information at the expense of PSNR.
The encoding performance depends on the number of bands and the setting parameters, but the proposed method is useful for MSFA-based imaging systems.


\section*{Acknowledgment}
This work was supported by JSPS KAKENHI Grant Number 15K20899. The authors would like to thank Yukio Kosugi from Tokyo Institute of Technology for providing the test images.

\ifCLASSOPTIONcaptionsoff
  \newpage
\fi


\end{document}